\documentclass[aps,pra,twocolumn,amsmath,amssymb,nofootinbib,showpacs,superscriptaddress]{revtex4-1}
\usepackage[english]{babel}
\usepackage{latexsym}
\usepackage{graphics}
\usepackage{graphicx}
\usepackage{epsfig}
\usepackage{color}
\usepackage{bm}
\usepackage{amsmath}
\usepackage{amssymb}
\usepackage{amsthm}
\usepackage{dcolumn}
\usepackage{bm}
\usepackage{float}
\usepackage{hyperref}
\usepackage{color}
\usepackage{epstopdf}
\usepackage{cleveref}
\usepackage[svgnames]{xcolor}
\usepackage{mathrsfs}
\hypersetup{hidelinks,colorlinks=true,allcolors=DarkBlue}

\usepackage{qcircuit}

\usepackage[toc,page]{appendix}

\begin{document}
	
	\newcommand{\ket}[1]{|#1\rangle} 
	\newcommand{\bra}[1]{\langle #1|}
	\newcommand{\bS}{\text{$\mathbf{S}$}}
	\newcommand{\braket}[2]{\langle #1|#2\rangle}
	
	\newcommand{\inp}{{\rm in}}
	\newcommand{\out}{{\rm out}}
	
	\newcommand{\init}{{\textrm{init}}}
	
	\newcommand{\kett}[1]{|#1\rangle\rangle}
	\newcommand{\braa}[1]{\langle\langle #1|}
	\newcommand{\com}[1]{{\color{blue}#1}}
	
	\newcommand{\Set}[2]{
		\left\{\, #1 \: \middle| \: #2 \, \right\}
	}
	
	\newcommand{\Abs}[1]{\left| #1 \right|}
	
	\newcommand{\Norm}[1]{\left\lVert #1 \right\rVert}
	
	\newcommand{\InnerProduct}[2]{\langle #1 , #2 \rangle}
	
	\newcommand{\Commutator}[2]{ \left[ #1 , \, #2 \right] }
	
	\newcommand{\AntiCommutator}[2]{ \left\{ #1 , \, #2 \right\} }
	
	\newcommand{\Real}{\mathbb{R}}
	\newcommand{\Comp}{\mathbb{C}}
	
	\newcommand{\Hil}{\mathcal{H}}
	
	\newcommand{\Bound}[1]{\mathfrak{B} \left( #1 \right)}
	
	\newcommand{\TrClass}[1]{\mathfrak{T} \left( #1 \right)}
	
	\newcommand{\States}[1]{\mathfrak{S} \left( #1 \right)}
	
	\newcommand{\Effects}[1]{\mathfrak{E} \left( #1 \right)}
	
	\newcommand{\StatesPure}[1]{\mathfrak{S}_{\rm pure} \left( #1 \right)}

	\newcommand{\Mat}[3]{\operatorname{Mat}_{#1 \times #2} \left( #3 \right)}
	
	\newcommand{\Tr}[1]{\,\operatorname{Tr} \left( #1 \right)}
	\newcommand{\TrPart}[2]{ \operatorname{Tr}_{#1} \left( #2 \right)}
	
	\newcommand{\Distributions}[1]{\mathfrak{P} \left( #1 \right)}
	\newcommand{\DistributionsPure}[1]{\mathfrak{P}_{\rm pure} \left( #1 \right)}
	\newcommand{\ChannelsCat}{\textbf{Channels}}
	\newcommand{\PStochCat}{\textbf{PStoch}}
	\newcommand{\TrPresCat}{\textbf{TrPres}}
	
	\newcommand{\QFunct}{\mathbf{Q}}
	\newcommand{\GFunct}{\mathbf{G}}
	
	\newcommand{\Hbf}{\textbf{H}}
	\newcommand{\ebf}{\textbf{e}}
	\newcommand{\Ebf}{\textbf{E}}
	\newcommand{\Dbf}{\textbf{D}}
	\newcommand{\Lbf}{\textbf{L}}
	
	\newtheorem{claim}{Claim}
	\newtheorem{Example}{Example}
	
	\preprint{APS/123-QED}
	
	\title{Minimal informationally complete measurements \\ for probability representation of quantum dynamics}
	
	\author{V.I. Yashin}
	\affiliation{Russian Quantum Center, Skolkovo, Moscow 143025, Russia}
	\affiliation{Moscow Institute of Physics and Technology, Dolgoprudny 141700, Russia}
	\affiliation{Steklov Mathematical Institute of Russian Academy of Sciences, Moscow 119991, Russia}
	
	\author{E.O. Kiktenko}
	\affiliation{Russian Quantum Center, Skolkovo, Moscow 143025, Russia}
	\affiliation{Moscow Institute of Physics and Technology, Dolgoprudny 141700, Russia}
	\affiliation{Steklov Mathematical Institute of Russian Academy of Sciences, Moscow 119991, Russia}
	
	\author{A.S. Mastiukova}
	\affiliation{Russian Quantum Center, Skolkovo, Moscow 143025, Russia}
	\affiliation{Moscow Institute of Physics and Technology, Dolgoprudny 141700, Russia}
	
	\author{A.K. Fedorov}
	\affiliation{Russian Quantum Center, Skolkovo, Moscow 143025, Russia}
	\affiliation{Moscow Institute of Physics and Technology, Dolgoprudny 141700, Russia}
	
	\date{\today}
	\begin{abstract}
		In the present work, we suggest an approach for describing dynamics of finite-dimensional quantum systems in terms of pseudostochastic maps acting on probability distributions, 
		which are obtained via minimal informationally complete quantum measurements.
		The suggested method for probability representation of quantum dynamics preserves the tensor product structure, which makes it favourable for the analysis of multi-qubit systems.
		A key advantage of the suggested approach is that minimal informationally complete positive operator-valued measures (MIC-POVMs) are easier to construct in comparison with their symmetric versions (SIC-POVMs).
		We establish a correspondence between the standard quantum-mechanical formalism and the MIC-POVM-based probability formalism.
		Within the latter approach, we derive equations for the unitary von-Neumann evolution and the Markovian dissipative evolution, which is governed by the Gorini-Kossakowski-Sudarshan-Lindblad (GKSL) generator.
		We apply the MIC-POVM-based probability representation to the digital quantum computing model. 
		In particular, for the case of spin-$1/2$ evolution, we demonstrate identifying a transition of a dissipative quantum dynamics to a completely classical-like stochastic dynamics.
		One of the most important findings is that the MIC-POVM-based probability representation 
		gives more strict requirements for revealing the non-classical character of dissipative quantum dynamics in comparison with the SIC-POVM-based approach.
		Our results give a physical interpretation of quantum computations and pave a way for exploring the resources of noisy intermediate-scale quantum (NISQ) devices.
	\end{abstract}
	
	\maketitle
	
	\section{Introduction}
	
	The problem of the description of quantum dynamics plays a significant role both in studying fundamental aspects of quantum physics~\cite{Schroeck1996} and exploring potential applications~\cite{Lukin2014}.
	In the latter case, it is crucial to highlight the role of non-classical phenomena and understand the origin of advantages of the use of quantum systems in various applications, such as quantum communication and quantum computing~\cite{NielsenChuang}. 
	This question is quite non-trivial, in particular, due to the fact that the commonly used descriptions of quantum states drastically differ from the language of statistical physics, which uses probability distributions.
	Several attempts to describe quantum systems using quantum analogues of probability distributions, such as the Wigner function~\cite{Wigner1932}, 
	have been made~\cite{Wigner1932,Glauber1963,Sudarshan1963,Glauber1969,Agarwal1970,Husimi1940,Kano1965,Ferrie2009}.
	Although the Wigner function cannot be interpreted as the probability distribution since it takes negative values, 
	its negativity can be linked to the resource providing quantum speed-up in solving computational problems~\cite{Galvao2006,Spekkens2008,Ferrie2011,Gottesman2012,Gottesman2014,Howard2014,Raussendorf2015,Pashayan2015}.
	Quasi-probability distributions of the other type such the Glauber-Sudarshan~\cite{Glauber1963,Sudarshan1963} and Husimi~\cite{Husimi1940} functions are also actively used for the description of quantum systems.
	Quantum phenomena can be also described on the language of tomographic distributions~\cite{Manko1996,Manko1997,Manko2010,Fedorov2013}, which are parametrized family of probability distributions.
	Quantum tomograms are related to Wigner functions via the Radon transformation~\cite{Lvovsky2009}. 
	Quantum tomography is essentially related to the question of the completeness of quantum measurements~\cite{Lvovsky2009,Busch1991}.
	
	Advances in understanding the role of various types of quantum measurements have formulated several new concepts. 
	In particular, quantum systems can be described via a probability distribution, which is obtained via informationally complete (IC) quantum measurements~\cite{Busch1991,Busch1995}.
	Since measurements in the quantum domain are represented by positive operator valued measures (POVMs), 
	the full determination of quantum states requires the use of so-called informationally complete POVMs (IC-POVMs)~\cite{Busch1991,Busch1995,Caves2004}.
	Importantly, this question is linked to the idea of using single measurement for quantum state characterization~\cite{Busch1995},
	and to the concept of the Husimi representation the quantum systems with continuous variables systems~\cite{Husimi1940}.
	We also note that probability structures behind quantum theory have been widely studied~\cite{Busch1991,Busch1995,Holevo2012}.

	An important special case of IC-POVM is its symmetric version, which is known as symmetric IC-POVMs (SIC-POVMs), where all pairwise inner products between the POVM elements are equal.
	SIC-POVMs are explored in various applications including tomographic measurements~\cite{Manko2010,Caves2002}, quantum cryptography~\cite{Fuchs2003}, and measurement-based quantum computing~\cite{Jozsa205}. 
	In addition, the idea of SIC-POVMs is actively used in quantum Bayesianism reformulation of quantum mechanics~\cite{Caves2002,Caves20022,Caves2004,Fuchs2013}.
	The quantum part of a classical probability simplex, which is achievable by measurements obtained via SIC-POVM (SIC-POVM measurements), is referred to as a qplex (i.e. a `quantum simplex')~\cite{Appleby2017}.
	
	The SIC-POVM formalism can be further extended for the description of dynamics of finite-dimensional quantum systems.
	The main difference with the quasi-probability representation is that quantum systems are described via (positive and normalized) probability distributions, which are obtained by SIC-POVMs~\cite{Kiktenko2020}.
	These probability distributions evolve under the action of pseudostochastic maps --- stochastic maps, which are described by matrices that may have {\it negative} elements.
	This idea, in a sense, changes the paradigm of revealing the distinction between quantum and classical dynamics.  
	Indeed it allows linking `quantumness' with negative probabilities can be extended to the study of non-classical properties of quantum dynamics and measurement processes~\cite{Wetering2017}. 
	Quantum dynamical equations both for unitary evolution of the density matrix governed by the von Neumann equation and dissipative evolution governed by Markovian master equation, 
	which is governed by the Gorini-Kossakowski-Sudarshan-Lindblad (GKSL) generator, can be derived~\cite{Kiktenko2020}. 
	Moreover, practical measures of non-Markovianity of quantum processes can be obtained and applied for studying existing quantum computing devices.
	However, this approach has a number of challenging aspects, which are related in particular to the problem of SIC-POVM existence, which is considered analytically and numerically just for a number of cases (for a review, see Ref.~\cite{Fuchs2017}).
	Thus, this representation is based on probability distributions, which are given by SIC-POVM measurements, is hardly applicable to the analysis of multi-qubit systems with an arbitrary number of qubits.
	This limits applications of such an approach, in particular, for the analysis of quantum information processing devices.

	\begin{figure}
		\includegraphics[width=1.0\linewidth]{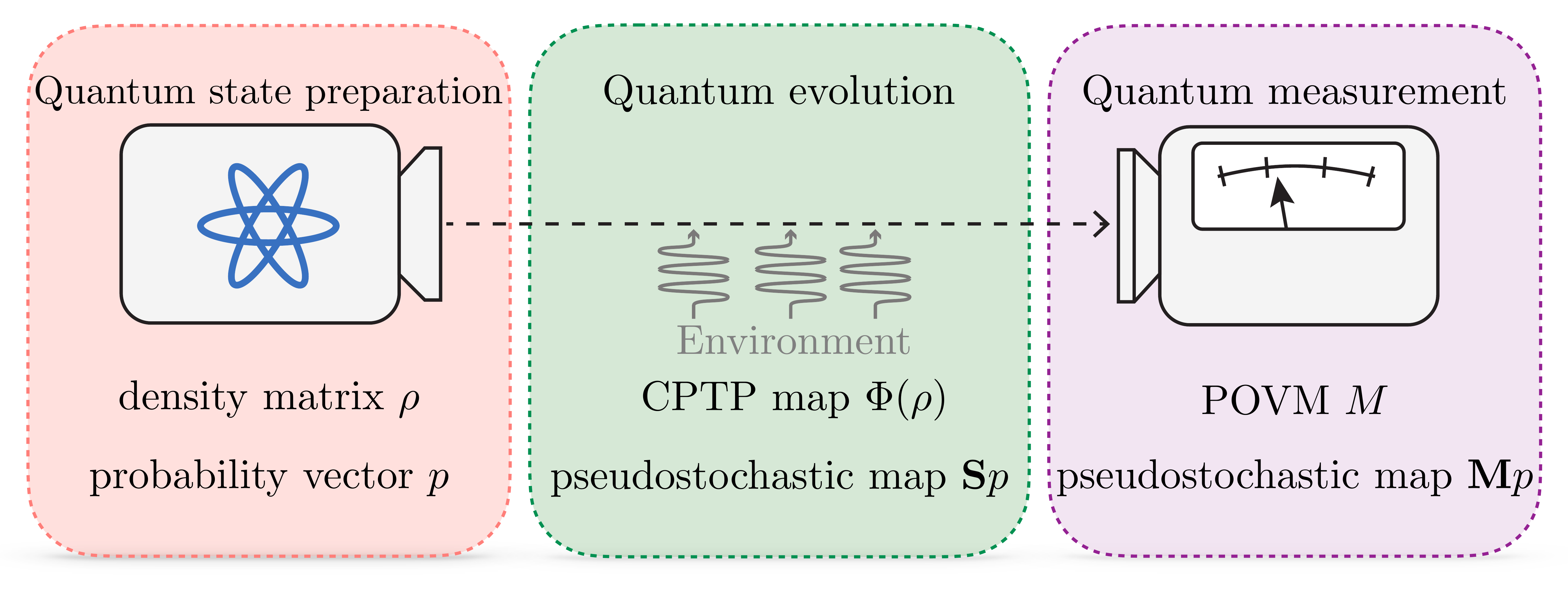}
		\vskip -2mm
		\caption{Quantum dynamics in terms of the commonly accepted density matrix approach (a quantum channel $\Phi(\cdot)$ acts on a density matrix $\rho$, and the measurement process is described by POVM $M$) and 
		the probabilistic description via MIC-POVM (a pseudostochastic map $\mathbf{S}$ acts on a MIC-POVM probability distribution $p$, and the measurement process is described by a pseudostochastic map $\mathbf{M}$).}
		\label{fig:sch}
	\end{figure}
	
	In this work, we present a generalization of the probability representation of quantum dynamics using minimal informationally complete positive operator-valued measures based (MIC-POVM) measurements, 
	which are an important class of quantum measurements~\cite{Weigert2000,Weigert2006,DeBrota2020,Smania2020,Planat2018}.
	For a $d$-dimensional Hilbert space an IC-POVM is said to be MIC-POVM if it contains exactly $d^2$ linearly independent elements. 
	Here we construct pseudostochastic maps that act on probability distributions, which are obtained by MIC-POVM measurements.
	We demonstrate that this approach is a generalization of the SIC-POVM-based representation~\cite{Kiktenko2020}, and it has a number of important features. 
	First, such an approach allows for preserving the tensor product structure, which is important for the description of multi-qubit systems.
	Second, MIC-POVMs are easier to construct in comparison with SIC-POVMs.
	Using the MIC-POVM-based probability representation, we derive quantum dynamical equations both for the unitary von-Neumann evolution and the Markovian dissipative evolution, 
	which is governed by the GKSL generator.
	It allows us to generalize previously obtained results on the description of the von-Neumann evolution of quantum systems on the probability language~\cite{Weigert2000}.
	We demonstrate how the suggested approach can be applied for the analysis of NISQ computing processes and obtain pseudostochastic maps for various single-qubit decoherence channels, as well as single-qubit and multi-qubit quantum gates.
	This gives an interpretation of quantum computations as actions of pseudostochastic maps on bitstring, 
	where the nature of quantum speedup is linked to the negative elements in pseudostochastic matrices that are corresponding to the quantum algorithm (as a sequence of gates and projective measurements).

	Our work is organized as follows.
	In Sec.~\ref{sec:construction}, we construct a probability representation via MIC-POVMs. 
	In Sec.~\ref{sec:dynamics}, we derive quantum dynamical equations in the MIC-POVM representation.
	In Sec.~\ref{sec:decoherence}, we use the probability representation to study a dissipative dynamics of a spin-1/2 particle.
	One of the key findings is that the MIC-POVM-based probability representation gives more strict requirements for revealing the non-classical character of dissipative quantum dynamics in comparison with the SIC-POVM-based approach. 
	In Sec.~\ref{sec:qcomp}, we demonstrate the applicability of the MIC-POVM-based probability representation for the analysis of quantum computing processes. 
	We illustrate our approach by considering Grover's algorithm.
	We summarize the main results and conclude in Sec.~\ref{sec:conclusion}.
	
	\section{Probability representation via MIC-POVMs}\label{sec:construction}
	
	Here we construct a MIC-POVM-based probability representation of quantum mechanics in the case of finite-dimensional systems.
	For this purpose, we first consider the representation of states, then study their evolution described by quantum channel, consider quantum measurements, and finally discuss a transition between representations defined by different MIC-POVMs.
	We also highlight here an important feature of the MIC-POVM probability representation, which is the simple tensor product structure. 
	The summary of results is presented in Table~\ref{tbl:1}.
		
	\begin{table*}[t]
		\begin{tabular}{p{0.17\linewidth}|p{0.34\linewidth}|p{0.36\linewidth}}
			& Standard formalism & MIC-POVM formalism \\ \hline \hline
			State & Density matrix $\rho\in \States{\Hil}$ & Probability vector $p \in \Distributions{E}$ \\
			Channel & CPTP map $\Phi: \States{\Hil^{\rm in}} \rightarrow \States{\Hil^{\rm out}} $ & Pseudostochastic matrix ${\bf S} \in \Mat{d^{2}}{d^{2}}{\mathbb{R}}$ \\
			Measurement & POVM $M=\{M_{i}\}_{i=1}^{m}$, $M_{i} \in \Effects{\Hil}$ & Pseudostochasitc matrix ${\bf M} \in \Mat{m}{d^{2}}{\mathbb{R}}$ \\\hline
			Tensor product rules & $\rho^{AB}=\rho^{A}\otimes\rho^{B}$ & $p^{AB}=p^{A} \otimes p^{B}$ \\
							  & $\Phi^{AB}=\Phi^{A}\otimes \Phi^{B}$ & ${\bf S}^{AB}={\bf S}^{A} \otimes {\bf S}^{B}$ \\
							  & $M^{AB}=M^{A}\otimes M^{B}=\{M^{A}_{i}\otimes M^{B}_{j}\}_{i=1,j=1}^{m_{1},m_{2}}$ & ${\bf M}^{AB} = {\bf M}^{A} \otimes {\bf M}^{B}$
		\end{tabular}
		\caption{The correspondence between standard and MIC-POVM-based formalisms.}
		\label{tbl:1}
	\end{table*}
	
	\subsection{Definitions and notations}\label{sec:preliminaries}
	
	We start our consideration by introducing basic definitions and notations. Let $\Hil$ be a $d$-dimensional Hilbert space, where $d$ is finite. 
	We introduce $\Bound{\Hil}$ as an algebra of bounded operators on $\Hil$.
	We also introduce the space of trace class operators $\TrClass{\Hil}$ and space of $n \times k$-matrices over the field $\mathbb{F}$, which we refer to as $\Mat{n}{k}{\mathbb{F}}$.
	Standardly, we use density operators $\rho \in \TrClass{\Hil}$ for the description of quantum states, where $\rho\geq0$ and $\Tr{\rho} = 1$.
	The convex space of quantum states is denoted as $ \States{\Hil} $. 
	Extreme points of this space are called \emph{pure states}, and we denote them as $\StatesPure{\Hil} $.
	An operator $B \in \Bound{\Hil}$ is called an \emph{effect}, if $0\leq{B}\leq {\bf I}_{d}$, where we use ${\bf I}_{n}$ to denote $n$-dimensional identity operator.
	The space of all effects is denoted by $\Effects{\Hil}$.
	The \emph{channel} $\Phi :\States{\Hil^{\rm in}}\rightarrow\States{\Hil^{\rm out}} $ is a trace-preserving, completely-positive (CPTP) linear map between states on Hilbert spaces $\Hil^{\rm in}$ and $\Hil^{\rm out}$.
	
	A set of effects $E=\{E_1,\ldots,E_m\}$ with $E_k\geq 0$ and that satisfies the condition $\sum_k E_k = {\bf I}_{d} $, is known as POVM (positive operator-valued measure). 
	The Born rule implies that a given state $\rho$ defines a probability distribution which we treat as a column-vector
	\begin{equation}
		p=\begin{bmatrix} p_1 & \ldots & p_m \end{bmatrix}^\top, \quad p_k=\Tr{\rho E_k},
	\end{equation}
	where $\top$ the denotes a standard transposition.
	The POVM $E$ is the MIC-POVM if it forms the basis of $ \Bound{\Hil} $. 
	In this case, $E$ contains $d^2$ elements, and every state $\rho$ is fully described by the corresponding probability vector $p$. 
	We denote a set of possible probability vectors with fixed MIC-POVM $E$ and varying $\rho$ as $\Distributions{E} $. 
	We note that SIC-POVMs are a particular class of MIC-POVMs. 
	
	The elements of SIC-POVM $E^{\rm sym}$ have the following form:
	\begin{equation}
		E_k^{\rm sym} = \frac{1}{d} \ket{\psi^{\rm sym}_k}\bra{\psi^{\rm sym}_k}, \quad k = 1,\ldots,d^{2}. 
	\end{equation}
	Here vectors $\ket{\psi^{\rm sym}_k}$ satisfy the following condition:
	\begin{equation} 
		|\langle{\psi^{\rm sym}_k}|\psi^{\rm sym}_l\rangle|^2 = \frac{d\delta_{kl}+1}{d+1},
	\end{equation}
	and $\delta_{kl}$ is the Kronecker symbol.
	
	As it is noted above, MIC-POVM and SIC-POVM measurements give probability distributions that fully describe quantum states. 
	Therefore, in order to describe the dynamics of quantum systems, one has to find corresponding maps acting on these probability distributions.
	We note that stochastic maps are not sufficient for the description of quantum dynamics.
	As it is shown in Ref.~\cite{Wetering2017}, corresponding maps for a description of dynamics of quantum states, which are presented by a probability distribution, are quasistochastic or pseudostochastic.
	In line with Ref.~\cite{Chruscinski2013, Chruscinski2015, Kiktenko2020} in our work we prefer to use a term pseudostochastic to emphasize that we deal with classical probability distributions rather than quasi-probabilities.
	
	We remind that a stochastic matrix $M\in\Mat{n}{k}{\Real} $ is a matrix, for which $ M_{ij} \geq 0, \sum_{i=1}^{n} M_{ij} = 1 $ for every $j=1,\ldots,k$. 
	It is called \emph{bistochastic}, if $n=k$ and also $\sum_j M_{ij} = 1$ for every  $i = 1,\ldots,n$.
	Under bistochastic map, a fully chaotic state $\begin{bmatrix} 1/n & \ldots & 1/n \end{bmatrix}$ remains the same. 
	For the description of quantum dynamics it is necessary to introduce pseudostochastic maps which can be presented as a matrix $M \in \Mat{n}{k}{\Real}$ with $\sum_{i=1}^{n} M_{ij} = 1$ but without the restriction on the positivity of matrix elements.
	The square $n\times n$ matrix is pseudobistochastic if for any $j\in\{1,\ldots n\}$ one has $\sum_{j=1}^{n} M_{ij} = 1$ as well  (again, some elements of $M_{ij} $ may be negative).
	
	\subsection{Representation of states}\label{sec:states}
	
	We consider a MIC-POVM $E=\{E_k\}_{k=1}^{d^2}$ in the $d$-dimensional Hilbert space $\Hil$.
	There is a canonical duality between spaces $\Bound{\Hil}$ and $\TrClass{\Hil}$, which is given by the bilinear form $(B,\rho) \mapsto \Tr{\rho B}$ for $\rho\in\TrClass{\Hil}$ and $B\in\Bound{\Hil}$. 
	It means that any linear functional on $\Bound{\Hil}$ can be represented as $\Tr{\rho~\cdot}$, and any functional on $\TrClass{\Hil}$ can be represented as $\Tr{\cdot~B}$.
	
	Since MIC-POVM $E$ forms a linear basis in $\Bound{\Hil}$ one can construct a basis $e = \{e_i\}_{i=1}^{d^{2}} $ in $\TrClass{\Hil}$, such that $\Tr{E_l e_k} = \delta_{l,k}$.
	This basis is usually referred to as a \emph{dual} basis to $E$. 
	Explicitly, the elements of this basis are as follows:
	\begin{equation}
		e_l = \sum_{k=1}^{d^2} ({\bf T}^{-1})_{lk} E_k , \quad {\bf T}_{nm} = \Tr{E_n E_m}.
	\end{equation}
	Then an arbitrary state $\rho \in \States{\Hil}$ can be represented in the following form:
	\begin{equation}\label{rho}
		\rho = \sum_{k=1}^{d^2}{p_k e_k}, \quad p_{k}={\rm Tr}(\rho E_{k}).
	\end{equation}
	
	We note that in the SIC-POVM case we have
	\begin{equation}
		\begin{aligned}
			{\bf T}_{nm} &= \frac{d\delta_{nm}+1}{d^{2}(d+1)}, \\
			({\bf T}^{-1})_{nm} &= d(d+1)\delta_{nm} - 1
		\end{aligned}
	\end{equation}
	
	Let  $s$ and $p$ be two probability vectors corresponding to density operators $\rho$ and $\sigma$.
	Then we can introduce a probability representation of the Hilbert--Schmidt product:
	\begin{equation}\label{eq:HSproduct}
		{\rm Tr} (\rho \sigma) = \sum_{n,m=1}^{d^{2}} p_{n} s_{m} {\rm Tr} (e_{n} e_{m}) = s^\top {\bf T}^{-1} p,
	\end{equation}
	and an analog of the matrix-matrix multiplication:
	\begin{multline}
		\left(s*p\right)_k \equiv \Tr{\sigma \rho E_k}\\
		=\sum_{n,m=1}^{d^{2}} \Tr{e_n e_m E_k} s_n p_m =  p^{\top}  \Lambda^{(k)} s.
	\end{multline}
	Here $\Lambda^{(k)}$ is a $d^{2} \times d^{2}$ matrix with elements $\Lambda^{(k)}_{nm} = \Tr{e_n e_m E_k}$.

\begin{figure}
		\includegraphics[width=0.7\linewidth]{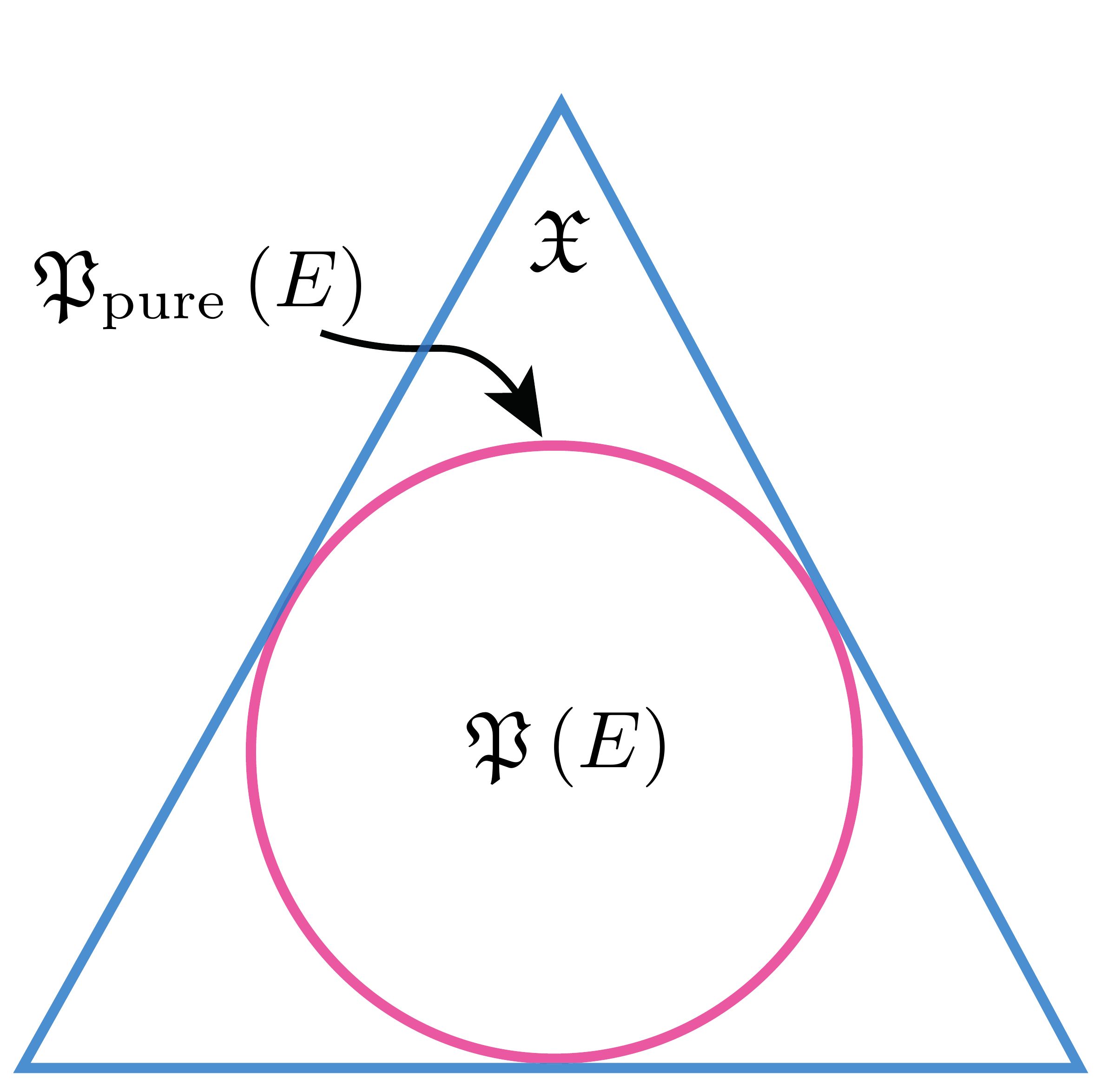}
		\vskip -4mm
		\caption{Schematic two-dimensional diagram showing relations, and points of contact to the probability simplex $\mathfrak{X}$ and qplex $\Distributions{E}$ with a border $\DistributionsPure{E}$. 
		Here the simplex is presented in the form of a triangle, since it has finite number of extreme points, while the qplex has an infinite number of extreme points and thus it is presented as a sphere.}
		\label{fig:Qplex}
	\end{figure}

	One can check that operators $e_k$ have a unit trace, but they are not necessarily positive. 
	Therefore, not any probability vector $p$ corresponds to a quantum state. 
	The set of possible distributions $\Distributions{E}$ has a form
	\begin{equation}\label{Distr}
		\Distributions{E} = \Set{p}{\sum_{k=1}^{d^2}{p_k}= 1, \: \sum_{k=1}^{d^2}{p_k e_k}\geq 0}.
	\end{equation}
	which is referred to as qplex (see Ref.~\cite{Appleby2017}).
	The set of distributions corresponding to pure states is as follows:
	\begin{equation}\label{DistrPure}
		\DistributionsPure{E}=\Set{p}{\sum_{k=1}^{d^2} p_k = 1, \: p * p = p}.
	\end{equation}
	The convex hull of this set is a set of distributions corresponding to all states $\Distributions{E}$. 
	We show schematic diagram of relations between sets $\Distributions{E}$, $\DistributionsPure{E}$, and the full $d^{2}$-dimensional simplex $\mathfrak{X}$ in Fig.~\ref{fig:Qplex}.
	
	Since $\Distributions{E}$ does not occupy the full space of $d^{2}$-dimensional simplex $\mathfrak{X}$ it is valuable to have a method for checking whether given distribution $p$ belongs to $ \Distributions{E}$.
	A straightforward way to cope with this task is to apply Eq.~\eqref{rho} to reconstruct $\rho$ and check whether $\rho\geq 0$.
	However, in the present work, we are interested in a method that does not require a transition to the standard formalism.

	Consider a characteristic polynomial of a density operator $\rho$
	\begin{equation}
		\chi(\lambda) = \prod_{k=1}^d (\lambda - \lambda_k),
	\end{equation}
	where $\{ \lambda_n \}_{n=1}^d $ is the spectrum of $\rho$. 
	Let us define a set $\{a_n\}_{n=1}^d$ with the following elements: 
	\begin{equation}
		a_n := \Tr{\rho^n} = \sum_l (p^{*n})_l = \sum_{k=1}^{d} \lambda_k^n, \quad n=1,\ldots,d.
	\end{equation}
	In order to check that $ p \in \Distributions{E} $, it is necessary and sufficient to check that $ \lambda_k \geq 0 $ for all $k$. 

	Using the Newton--Girard identities the characteristic polynomial can be rewritten in the form
	\begin{equation}
		\chi(\lambda) = \sum_{m=0}^d (-1)^m b_m \lambda^{d-m},
	\end{equation}
	where $b_{0}= 1$ and
	\begin{equation}
		b_n = \frac{1}{n}\sum_{i=1}^n (-1)^{i-1} b_{n-i} a_i, \quad n=1,\ldots,d.
	\end{equation}
	
	If starting from some $d'\in\{1,\ldots,d\}$
	\begin{equation}
	b_{d'}=b_{d'+1}=\ldots=b_{d}=0,
	\end{equation} 
	then $\chi(\lambda)$ has $d-d'+1$ zero roots.
	It is convenient to remove them from consideration by resetting
	\begin{equation}
		\chi(\lambda) := \chi(\lambda) / \lambda^{d-d'+1}.
	\end{equation}
	Otherwise we set $d':=d$.
	
	Then we suggest using the Routh--Hurwitz criterion in order to verify that every root of the polynomial $\chi(\lambda)$ is nonnegative. 
	Let
	\begin{equation}
		\widetilde{\chi}(\lambda) = \prod_{k=1}^{d'}(\lambda + \lambda_k) = (-1)^{d'} \chi(-\lambda) = \sum_{m=0}^{d'} b_m \lambda^{d'-m}.
	\end{equation}
	One can see that nonnegative roots of $\chi(\lambda)$ imply nonpositive roots of $\widetilde{\chi}(\lambda)$.
	The Routh--Hurwitz criterion states that every root of $\widetilde{\chi}(\lambda)$ is negative if and only if the principal minors $\{\Delta_i\}_{i=1}^{d'}$ of the Hurwitz matrix
	\begin{equation}
		\mathscr{H}=\begin{bmatrix}
			b_1 & b_3 & b_5 & \cdots  & 0 & 0\\
			b_0 & b_2 & b_4 & \cdots &  0 & 0\\
			0 & b_1 & b_3 & \cdots & 0 & 0 \\
			\vdots & \vdots & \vdots & \ddots & \vdots & \vdots \\
			0 & 0 & 0 & \ldots & b_{d'-2} & b_d'
		\end{bmatrix}
	\end{equation}
	are positive:
	\begin{equation}
		\Delta_1 > 0, \: \Delta_2 > 0, \: \Delta_3 > 0, \: \ldots , \: \Delta_{d'} > 0. 
	\end{equation}
	These relations form the constructive way for checking whether $p \in \Distributions{E}$.

	\subsection{Representation of tensor products}\label{Tensor}

	The use of MIC-POVM probability vectors allows one to employ a simpler description of tensor products.
	This is an important advantage in comparison to the SIC-POVM case~\cite{Kiktenko2020}.
	Let $ \Hil^A$ and $\Hil^B $ be Hilbert spaces with MIC-POVMs $E^A=\{E_i^A\}_{i=1}^{d_A^2}$ and $E^B=\{E_i^B\}_{i=1}^{d_B^2}$, correspondingly. 
	We then take MIC-POVM 
	\begin{equation}
		E^{AB}=\{E_i^A\otimes E_j^B\}_{i,j=1}^{d_A^2,d_B^2}\equiv E^A \otimes E^B 
	\end{equation}
	on the space $\Hil^A \otimes \Hil^B$.
	If $ e^A $ and $ e^B $ are dual bases for $ E^A$ and $E^B $, then $ e^{AB} = e^A \otimes e^B $ is dual to $ E^{AB} $. 
	If $ \rho^A \in \States{\Hil^A}, \rho^B \in \States{\Hil^B} $ are states and $p^A, p^B$ are corresponding probability vectors, then the probability vectors of $\rho^A \otimes \rho^B$ is as follows:
	\begin{equation}
		p^{AB}_{(kl)} = \Tr{ ( \rho^A \otimes \rho^B) (E^A_k \otimes E^B_l) } = p^A_k p^B_l,
	\end{equation}
	Here we use notation $(\alpha,\beta)$ with $\alpha\in\{1,\ldots,d_{A}\}$ and $\beta\in\{1,\ldots,d_{B}\}$ to define a multiindex.
	One can think that $(\alpha,\beta)\equiv(d_{A}-1)\alpha + \beta$ according the the standard Kronecker product rules.
	In the vector form, we have $p^{AB}=p^{A}\otimes p^{B}$.

	\subsection{Representation of channels}

	The next step is to obtain the probability representation of quantum channels.
	Let $\Hil^{\rm in}$, $\Hil^{\rm out}$ be Hilbert spaces with MIC-POVMs $E^{\rm in}, E^{\rm out}$ and $\Phi : \States{\Hil^{\rm in}} \rightarrow \States{\Hil^{\rm out}}$ be a quantum channel (CPTP map).  
	Consider a state $\rho^{\rm in} \in \States{\Hil^{\rm in}}$ and let $\rho^{\rm out}= \Phi(\rho^{\rm in})\in \States{\Hil^{\rm out}}$.
	Denote the probability vectors corresponding to $\rho^{\rm in}$ and $\rho^{\rm out}$ as $p^{\rm in}$ and $p^{\rm out}$ respectively. 
	Then the channel $\Phi$ can be characterized with a matrix $\bf{S}$ such that
	\begin{equation}
		p^{\rm out} = {\bf{S}} p^{\rm in}, \quad {\bf{S}}_{lk} =  \Tr{ E^{\rm out}_l \Phi(e_k^{\rm in}) }.
	\end{equation}
	This matrix is generally pseudostochastic (i.e. $\sum_k {\bf{S}}_{kl} = 1$), but it is not necessarily stochastic. 
	An action of the channel $\Phi$ on the state $\rho^{\rm in}$ can be written as follows:			
	\begin{equation}\label{eq:channel-from-matr}
		\Phi: \rho^{\rm in} \mapsto \sum_{k,l=1}^{d_{\rm out},d_{\rm in}} {\bf{S}}_{kl} e^{\rm out}_k \Tr{ E_l^{\rm in} \rho^{\rm in} }.
	\end{equation}
	In turn, an action of the dual channel $\Phi^{*}: \TrClass{\Hil^{\rm out}} \rightarrow \TrClass{\Hil^{\rm in}} $ is then given by
	\begin{equation}\label{eq:dual-channel-from-matr}
		\Phi^{*}: M^{\rm out} \mapsto \sum_{k,l=1}^{d_{\rm in},d_{\rm out}} {\bf{S}}_{kl} E^{\rm in}_k \Tr{ e_l^{\rm out} M^{\rm out} }.
	\end{equation}

	In the case of Kraus representation where the operation of the channel is defined in the form 
	\begin{equation}\label{eq:Kraus}
		\Phi(\rho){=}\sum_n V_n \rho V_n^{\dagger},
	\end{equation}
	the corresponding elements of the pseudostochastic matrix are
	\begin{equation}
		{\bf{S}}_{lk} = \sum_n \Tr{E^{\rm out}_l V_n e_k^{\rm in} V_n^{\dagger}}.
	\end{equation}
	
	The representation of tensor products for quantum channels can be used similarly to Sec.~\ref{Tensor}.
	If 
	\begin{equation}
		\Phi^A : \States{\Hil^{{\rm in},A}} \rightarrow \States{\Hil^{{\rm out},A}} 
	\end{equation}
	and 
	\begin{equation}
		\Phi^B : \States{\Hil^{{\rm in},B}} \rightarrow \States{\Hil^{{\rm out},B}} 
	\end{equation}
	are quantum channels with corresponding pseudostochastic matrices ${\bf{S}}^A$ and ${\bf{S}}^B$, then
	\begin{multline}
		{\bf{S}}^{AB}_{(kl)(nm)} = \\
		={\rm Tr} \left(
			\left\{
		 		E^{{\rm out},A}_k \otimes E^{{\rm out},B}_l 
		 	\right\}
			\left\{
				 \Phi^A \otimes \Phi^B 
		 	\right\}				 
			\left\{
				 e^{{\rm in},A}_{n} \otimes e^{{\rm in},B}_{m} 
			\right\}
		\right)
		\\=\Tr{ E^{{\rm out},A}_k \Phi^A(e_n^{{\rm in},A}) \otimes E^{{\rm out},B}_l \Phi^B(e_{m}^{{\rm in},B}) }\\ = {\bf{S}}^A_{(k,n)} \otimes {\bf{S}}^B_{(l,m)}.
	\end{multline}
	Thus, the tensor product of two quantum channels maps to the tensor product of two corresponding matrices.
	
	The channel of a partial trace
	\begin{equation}
		\TrPart{\mathcal{H}^{B}}{\cdot}:\rho^{AB} \mapsto \TrPart{\mathcal{H}^{B}}{\rho^{AB}} 
	\end{equation}
	taking an input $\rho^{AB} \in \States{\mathcal{H}^{A} \otimes \mathcal{H}^{B}}$
	then corresponds to the matrix in the form:
	\begin{multline}
		{\bf{S}}_{l(nm)} = \Tr{E^A_l \TrPart{B}{e^A_n \otimes e^B_m}} \\ 
		= \Tr{E^A_l e^A_n} = \delta_{ln}.
	\end{multline}
		
	As in the case of states, not any pseudostochastic matrix ${\bf{S}}$ corresponds to a (physical) quantum channel $\Phi$.
	In order to formulate a criterion, we use the Choi--Jamio{\l}kowski duality~\cite{Jamiolkowski1972,Choi1975}. 
		
	Let $ \Phi : \States{\Hil^{\rm in}} \rightarrow \States{\Hil^{\rm out}} $ be a trace-preserving map. 
	By fixing the orthonormal basis $\{\ket{n}\}_{n=1}^{d_{\rm in}}$ in $\Hil^{\rm in}$ ($d_{\rm in}={\rm dim} \Hil^{\rm in}$), we define a state $\sigma \in \States{\Hil'^{\rm in} \otimes \Hil^{\rm in}}$ with $\Hil'^{\rm in}=\Hil^{\rm in}$ of the following form:
	\begin{equation} \label{eq:max-ent-state}
		\sigma=\frac{1}{d_{\rm in}} \sum_{n,m=1}^{d_{\rm in}} \ket{n}\bra{m} \otimes \ket{n}\bra{m}.
	\end{equation}
	One can see that it is a density matrix of the pure state 
	\begin{equation}
		\ket{\phi}= \frac{1}{\sqrt{d_{\rm in}}}\sum_{k=1}^{d_{\rm in}} \ket{k} \otimes \ket{k}. 
	\end {equation}
	
	We then call \emph{Choi state} an operator $ \rho_\Phi $
	\begin{equation} 
		\rho_\Phi = ({\rm Id} \otimes \Phi) (\sigma) = \frac{1}{d_{\rm in}} \sum_{n,m=1}^{d_{\rm in}} \ket{n}\bra{m} \otimes \Phi(\ket{n}\bra{m}),
	\end{equation}
	where ${\rm Id}$ is an identical map.
	The Choi--Jamio{\l}kowski isomorphism says that the operator $ \rho_\Phi $ is a quantum state if and only if $\Phi$ is a quantum channel. 
	One can reconstruct an action of $\Phi$ on an arbitrary input using the following formula:
	\begin{equation}\label{eq:purestate-for-Choi}
		\Phi(\rho) = \TrPart{\Hil'^{\rm in}}{ ( \rho^{\rm in\top} \otimes {\bf I}_{d_{\rm in}}) \rho_\Phi}.
	\end{equation}
	
	The Choi--Jamio{\l}kowski isomorphism can be naturally formulated in the probability representation.
	Let $ {\bf{S}} $ be a matrix corresponding to a trace-preserving map $ \Phi $, and $ s $ be a vector corresponding to $\sigma$.
	We assume that $s$ is obtained with MIC-POVM $E^{\rm in} \otimes E^{\rm in}$.
	Then the \emph{Choi} probability vector has the form
	\begin{equation}\label{eq:Choi-prob}
		p_{\bf{S}} = \left( {\bf I}_{d_{\rm in}^{2}} \otimes {\bf{S}} \right) s
	\end{equation}
	One can see that $ {\bf{S}} $ corresponds to the quantum channel only in case $ p_{\bf{S}} \in \Distributions{E^{\rm in} \otimes E^{\rm out}} $. 
	In order to reconstruct $ {\bf{S}} $ via the vector $p_{\bf{S}}$, one can use the following relation:
	\begin{multline}
		{\bf{S}}_{lk} = \Tr{E^{\rm out}_l \Phi(e_k^{\rm in})} = \Tr{\rho_\Phi (e^{\rm in \top}_k \otimes E^{\rm out}_l)} \\
		= \sum_{n,m=1}^{d_{\rm in}^{2}} p_{{\bf S}{(nm)}} \Tr{ e_n^{\rm in} e^{\rm in \top}_k } \Tr{ e^{\rm out}_m E^{\rm out}_l} \\
		= \sum_{n=1}^{d_{\rm in}^{2}}  \Tr{ e^{\rm in}_n e^{\rm in \top}_k } p_{{\bf S}{(n,l)}}.
	\end{multline}
	
	It is useful to define $s$ in terms of probability vectors without the notion of the Hilbert space $\Hil^{\rm in}$ and the state $\sigma$. 
	Consider random pure state $\ket{\psi_{1}}\bra{\psi_{1}} \in \States{\Hil^{\rm in}}$ and denote its probability vector as $p^{(11)}$.
	Let us construct as a set of orthonormal probability vectors $\{p^{(kk)}\}_{k=1}^{d_{\rm in}}$ using the following equations for each $k = 2,\ldots,d_{\rm in}$:
	\begin{equation}\label{eq:set-for-orthonormal-set}
		\begin{aligned}
			&\sum_{l=1}^{d_{\rm in}} p^{(kk)}_l = 1; \\
			&p^{(kk)}*p^{(kk)} = p^{(kk)};\\ 
			&\sum_{l=1}^{d_{\rm in}} (p^{(kk)} * p^{(nn)})_l = 0, \quad n = 1,\ldots,k-1,
		\end{aligned}
	\end{equation}
	where we consider orthonormality with respect to the Hilbert-Schmidt product~\eqref{eq:HSproduct}.
	One can think about $p^{(kk)}$ as a probability vector of state $\ket{\psi_{k}}\bra{\psi_{k}}$ taken from an orthonormal basis constructed from $\ket{\psi_{1}}\bra{\psi_{1}}$.
	Of course, Eq.~\eqref{eq:set-for-orthonormal-set} has infinite number of solutions.
	
	Then the vectors $p^{(nm)}$ with $n\neq m$, corresponding to states $\ket{\psi_{n}}\bra{\psi_{m}}$, can be obtained using straightforward multiplicative relations. 
	For example, vector $ p^{(12)} $ can be obtained as a solution of the following equations:
	\begin{equation}
		\begin{aligned}
			p^{(12)} * p^{(kk)} &= 0,~~k\neq2; \quad &p^{(kk)} * p^{(12)} &= 0,~~k\neq1; \\
			p^{(12)} * p^{(11)} &= 0; \quad  &p^{(11)} * p^{(12)} &= p^{12}; \\
			p^{(12)} * p^{(12)} &= 0;  \quad  &p^{(12)} * p^{(12)} &= 0; \\
			p^{(12)} * p^{(22)} &= p^{12};  \quad  &p^{(22)} * p^{(12)} &= 0. \\
		\end{aligned}
	\end{equation}
	By finding $ p^{(nm)} $ for all $n,m=1,\ldots,d_{\rm in}$, we obtain the Choi distribution in the form 
	\begin{equation}
		p_{\bf{S}} = \frac{1}{d_{\rm in}} \sum_{n,m=1}^{d_{\rm in}} p^{(nm)} \otimes {\bf{S}} p^{(nm)}.
	\end{equation}

	We also would like to mention a special case where $E^{\rm in}=\{\ket{\psi_{i}^{\rm sym}}\bra{\psi_{i}^{\rm sym}}\}_{i=1}^{d_{\rm in}}$ is a SIC-POVM.
	Let $\overline{E}^{\rm in}=\{\ket{\overline{\psi}_{i}^{\rm sym}}\bra{\overline{\psi}_{i}^{\rm sym}}\}_{i=1}^{d_{\rm in}}$ with 
	\begin{equation}
		\ket{\overline{\psi}_{i}^{\rm sym}} = \sum_{i=1}^{d_{\rm in}} \ket{i} \overline{\bra{i} \psi^{\rm sym}_{i} \rangle} = \sum_{i=1}^{d_{\rm in}} \ket{i} \bra{\psi^{\rm sym}_{i} }i \rangle,
	\end{equation}
	where $\overline{x}$ stands for complex conjugate of $x$, and $\{\ket{n}\}_{n=1}^{d_{\rm in}}$ is a computational basis as usual.
	Then the probability vector of the state $\sigma = \ket{\phi}\bra{\phi}$ (see Eq.~\eqref{eq:purestate-for-Choi}) takes the following form with respect to the MIC-POVM $\overline{E}^{\rm in} \otimes E^{\rm in}$:
	\begin{multline}
		s_{(nm)} = \Tr{ \ket{\phi} \bra{\phi} (\overline{E}^{\rm in}_n \otimes E^{\rm in}_m) } = \\
		= \frac{1}{d_{\rm in}^2} \sum_{k,l}  \langle k\ket{\overline{\psi}^{\rm sym}_n} {\bra{\overline{\psi}^{\rm sym}_n}} l \rangle \langle k \ket{\psi^{\rm sym}_m}\bra{\psi^{\rm sym}_m} l \rangle = \\
		= \frac{1}{d_{\rm in}} \Abs{ \langle{\psi_m}|\psi_n\rangle }^2 = \frac{d_{\rm in} \delta_{nm} + 1}{d_{\rm in}^2 (d_{\rm in}+1)}.
	\end{multline}
	It then can be substituted to Eq.~\eqref{eq:Choi-prob} in order to obtain a Choi probability vector and verify that it corresponds to valid quantum state.
	
	\subsection{Representation of measurements}
	
	Here we consider a MIC-POVM-based probability representation of an arbitrary measurement with the finite number of outcomes.
	In the general case it is given by a POVM $M=\{M_{i}\}_{i=1}^{m}$ with some finite $m$.
	Note that $M$ may not belong to MIC class.
	According to the Born rule the probability to obtain $i$th outcome for an input state $\rho$ is given by
	\begin{equation}
		q_{i} = \Tr{\rho M_{i}}, \quad i=1,\ldots,m
	\end{equation}
	(here we assume that $M$ and $\rho$ are defined with respect to the same $d$-dimensional Hilbert space $\Hil$).
	Taking $\rho$ in the probability representation given by Eq.~\eqref{rho}, we obtain the following expression for the probability vector:
	\begin{equation}
	\begin{split}
		q&=\begin{bmatrix}q_{1} & \ldots & q_{m}\end{bmatrix}^{\top}, \\
		q&= {\bf M}p, \quad {\bf M}_{ij} = \Tr{M_{i}e_{j}}.
	\end{split}
	\end{equation}
	One can see that ${\bf M}$ is $m\times d^{2}$ pseudostochastic matrix because of normalization condition $\sum_{i=1}^{m}M_{i}={\bf I}_{d}$.
	We note that given matrix ${\bf M}$, the effects of the POVM in the standard formalism are given by
	\begin{equation}
		M_k = \sum_{l=1}^{d^{2}} \Tr{M_k e_l} E_l = \sum_{l=1}^{d^{2}} {\bf M}_{kl} E_l.
	\end{equation}
	
	Next, we consider a problem of the verification that a given $m\times d^{2}$ pseudostochastic matrix ${\bf M}$ corresponds to some valid POVM $M$ with $m$ outcomes.
	An idea behind such a test is very similar to the case of states, which is considered in Sec.~\ref{sec:states}, with the main difference that we swap the basis $E$ and the dual basis $e$.

	Consider two operators $X,Y \in \TrClass{\Hil}$.
	Using a dual basis $\{e_{i}\}_{i=1}^{d}$ one can represent them with row-vectors 
	$\lambda=\begin{bmatrix} \lambda_{1} & \ldots \lambda_{d} \end{bmatrix}$ and $\mu = \begin{bmatrix} \mu_{1} & \ldots \mu_{d} \end{bmatrix}$ according to the following expressions:
	\begin{equation}
		\begin{aligned}
			X &= \sum_{i=1}^{d^{2}}\lambda_{i} E_{i}, \quad &\lambda_{i} &= \Tr{ e_{i} X}, \quad i=1,\ldots,d^{2} ;\\
			Y &= \sum_{j=1}^{d^{2}}\mu_{j} E_{j} \quad &\mu_{j} &= \Tr{ e_{j} Y},  \quad j=1,\ldots,d^{2}.
		\end{aligned}
	\end{equation}
	Note that the trace operation takes the form:
	\begin{equation}
		\Tr{X} = \sum_{i=1}^{d^{2}} \lambda_{i} \Tr{E_{i}} = \lambda \kappa
	\end{equation}
	with $\kappa = \begin{bmatrix} \Tr{E_{1}} & \ldots & \Tr{E_{d^{2}}} \end{bmatrix}^{\top}$.

	Then we can introduce a `multiplication' of vectors $\mu$ and $\lambda$, denoted by $\circledast$, as follows:
	\begin{multline}
		(\lambda \circledast \mu)_{k} \equiv \Tr{XY e_{k}} \\
		=\sum_{n,m=1}^{d^{2}} \Tr{E_n E_m e_k} \lambda_n \mu_m =  \lambda  \widetilde{\Lambda}^{(k)} \mu^{\top},
	\end{multline}
	where $ \widetilde{\Lambda}^{(k)}$ is $d^{2} \times d^{2}$ matrix with elements 
	\begin{equation}\label{eq:lambda-tilde}
		\widetilde{\Lambda}^{(k)}_{nm} = \Tr{E_n E_m e_k}.
	\end{equation}
	
	Now we are ready to describe the verification algorithm.
	The normalization condition  $\sum_{i=1}^{m}M_{i}={\bf I}_{d}$ follows from the fact that ${\bf M}$ is pseudostochastic.
	So the only remaining issue is to check the semi-positivity condition $M_{i} \geq 0$.
	We note that in the case of states we derived expressions for $\Tr{\rho}$, \ldots, $\Tr{\rho^{d}}$ from the probability representation of $\rho$ and then substitute them into Routh--Hurwitz-like criterion.
	Here we act in a similar manner.
	For each $i$th row of the matrix ${\bf M}$ set $\lambda^{(i)}:=\begin{bmatrix} {\bf M}_{i,1} & \ldots & {\bf M}_{i,d^{2}} \end{bmatrix}$ and compute
	\begin{equation}
		a^{(i)}_{j} := \lambda^{(i) \circledast j} \kappa =  \Tr{M_{i}^{j}}.
	\end{equation}
	Then proceed with same steps as in the case of states replacing ${\{a_{n}\}_{n=1}^{d}}$ with ${\{a_{i}\}_{n=1}^{d}}$.
	If the positivity condition is fulfilled for all $i=1,\ldots,m$, then ${\bf S}$ corresponds to  valid `physical' quantum measurement. 
	
	Finally, we consider the case a measurement given by some Hermitian operator $O=O^{\dagger}\in \TrClass{\Hil}$, also known as an \emph{observable}.
	To obtain its probability representation we first take it spectral decomposition in the form
	\begin{equation}
		O = \sum_{i=1}^{m} x_{i} \Pi_{i},
	\end{equation}
	where $\{x_{i}\}_{i=1}^{m}$ are physical quantities which can be observed, and $\{\Pi_{i}\}_{i=1}^{m}$ are complete set of orthogonal (self-adjoint) projectors: $\sum_{i=1}^{m}\Pi_{i} = {\bf I}_{d}$, $\Pi_{i}\Pi_{j}=\delta_{ij}$.
	One can consider a POVM $M_{O}$ with effects $\{\Pi_{i}\}_{i=1}^{m}$ and its corresponding pseudostochastic matrix ${\bf M}$ keeping in mind that each $i$th outcome corresponds the quantity $x_{i}$.
	However, it is important to note if one is interested in mean value for some state $\rho$ is given by $\langle O \rangle = \Tr{O \rho}$, then one can consider a row-vector
	\begin{equation}
		{\bf O}^{\rm mean} := \begin{bmatrix} x_{1} & \ldots & x_{m} \end{bmatrix}{\bf M} 
	\end{equation}
	and compute mean value as $\langle O \rangle = {\bf O}_{\rm mean} p$, where $p$ is a probability vector of $\rho$.
	
	Finally, we note that the rules of the tensor product remain the same as in the case of quantum channels: 
	Pseudostochastic matrix of measurements on several physical subsystems is given by a tensor product of pseudostochastic measurements on each of subsystems.
	
	\subsection{Transitions between MIC-POVM-based representations}
	
	Up to this point the MIC-POVM, which determines the probability representation, was fixed.
	Here we consider a question of how to make a transition between representations determined by different MIC-POVMs.
	Let $E$ and $F$ be two MIC-POVMs defined with respect to the same $d$-dimensional Hilbert space $\Hil$, and let $e$ and $f$ be their corresponding dual bases.
	We use superscript $[E]$ or $[F]$ to emphasize that given probability vector or pseudostochastic matrix of a channel/measurement is given in $E$- or $F$-based representation.
	
	Consider elements of a pseudostochastic matrix of the measurement in $E$ given in $F$-based representation:
	\begin{equation}
		\left( {\bf M}_{E}^{[F]} \right)_{m,n} = \Tr{E_{m} f_{n}}, \quad m,n=1,\ldots,d^{2}.
	\end{equation}
	One can see the pseudostochastic matrix of the measurement in $F$ given in $E$-based representation is given by ${\bf M}_{F}^{[E]} = \left( {\bf M}_{E}^{[F]} \right)^{-1}$
	
	Then we come to the following relations:
	\begin{eqnarray}
		&p^{[E]} &= {\bf M}_{E}^{[F]} p^{[F]}, \\
		&{\bf S}^{[E]} &= {\bf M}_{E}^{[F]} {\bf S}^{[F]} {\bf M}_{F}^{[E]}, \\
		&{\bf M}^{[E]} &= {\bf M}^{[F]} {\bf M}^{[E]}_{F},
	\end{eqnarray}
	where $p^{[\cdot]}$, ${\bf S}^{[\cdot]}$, and ${\bf M}^{[\cdot]}$ corresponds to some state, channel, and POVM, correspondingly.
	
	\section{Dynamical equations}\label{sec:dynamics}
	
	Here we apply the developed MIC-POVM-based representation to quantum evolution equations (master equations).
	We consider two conceptually important cases.
	The first is the Liouville-von Neumann equation corresponding to the unitary evolution of quantum states.
	The second is the dissipative evolution governed by a Markovian master equation, which is governed by the GKSL generator.
	In the both cases we restrict ourself with a condition that generators are time-independent.
	
	\subsection{Liouville-von Neumann equation}
	
	Consider a $d$-dimensional dimensional Hilbert space $\Hil$.
	The evolution of a quantum state under the Hamiltonian $H=H^{\dagger} \in \TrClass{\Hil}$ is described by the Liouville-von Neumann equation
	\begin{equation}
		\dot{\rho}(t) = - \frac{\imath}{\hbar} \Commutator{H}{\rho(t)},
	\end{equation}
	where $ \Commutator{\cdot}{\cdot}$ denotes commutator. 
	In what follows we use dimensionless units and set $\hbar\equiv1$.
	
	Using Eq.~\eqref{rho}, then left multiplying by $E_l$ and taking trace, the equation takes the form
	\begin{equation}
		\dot{p}_l(t) = - \imath \sum_{k=1}^{d^{2}} \Tr{E_l \Commutator{H}{e_k}} p_k(t).
	\end{equation}
	
	Therefore, the Liouville-von Neumann equation takes the form of the ordinary linear differential equation with generator $\Hbf$
	\begin{equation}\label{eq:vNeq}
		\dot{p}(t) = \Hbf p(t), \quad \Hbf_{l,k} = \imath \Tr{H \Commutator{E_l}{e_k}}.
	\end{equation}
	We note that such a form of the equation for the unitary dynamics on the probability language has been obtained in Ref.~\cite{Weigert2000}. 
	In the following section we present the generalization of this results for the case of the GKSL equation.
	
	The $d^{2} \times d^{2}$ matrix $ \Hbf$ is a probabilistic representation of the Hamiltonian, which has following properties. 
	\begin{enumerate}
		\item The matrix $\Hbf$ is real: $\Hbf \in \Mat{d^{2}}{d^{2}}{\mathbb{R}}$.
		\item The sum of each column is zero: $\sum_{l=1}^{d^{2}} \Hbf_{l,k} = 0$.
	\end{enumerate}
	The first property follows from the fact that $\Commutator{E_{l}}{e_{k}}$ is Hermitian, and second fact come from the normalization condition on MIC-POVM effects $E_{i}$.
	It is worth to note that if $E$ is a SIC-POVM, then $\Hbf$ becomes antisymmetric ($\Hbf^{\top}=-\Hbf$), and thus all its diagonal elements are zero, and all rows also sum to zero (see Ref.~\cite{Kiktenko2020} for more details).
	
	The solution to Eq.~\eqref{eq:vNeq} can be presented in the form
	\begin{equation}\label{eq:unitary-stoch}
		p(t) = {\bf U}(t) p_{0}\quad {\bf U}(t)=e^{\Hbf t},
	\end{equation}
	where $p_{0}$ is a probability vector at $t=0$.
	Note that ${\bf U}(t)$ is pseudostochastic.
	The unitarity of the evolution governed by the Liouville-von Neumann equation implies preserving the Hilbert-Schmidt product between two arbitrary vectors during their evolution according to Eq.~\eqref{eq:vNeq}.
	Taking into account the probability representation of the Hilbert-Schmidt product is given by Eq.~\eqref{eq:HSproduct} we obtain the identity
	\begin{equation}
		{\bf T}^{-1} = {\bf U}(t)^{\top} {\bf T}^{-1} {\bf U}(t).
	\end{equation}
	Considering small times $t=\delta t \ll 1$ and expanding exponent of the evolution operator in Eq.~\eqref{eq:unitary-stoch} into Taylor series we arrive at the 3rd property of the MIC-POVM-based representation of a Hamiltonian.
	\begin{enumerate}
		\item[3.]	The following identity holds: 
		\begin{equation}\label{eq:extra-cond}
			\Hbf^{\top} {\bf T}^{-1} + {\bf T}^{-1} \Hbf =0.
		\end{equation}
	\end{enumerate}
	
	Consider a matrix $\widetilde{\Hbf}:={\bf T}^{-1} \Hbf$.
	One can see that since ${\bf T}$ is symmetric, $\widetilde{\Hbf}$ is antisymmetric: $\widetilde{\Hbf}^{\top} + \widetilde{\Hbf}=0$. 
	Combining this fact with the property 2 one has $\sum_{i=1}^{d^{2}}({\bf T} \widetilde{\Hbf})_{ij}=0$.
	Antisymmetric matrix $\widetilde{\Hbf}$ possessing these properties can be defined with $(d^{2}-2)(d^{2}-1)/2$ independent parameters.
	Note that for $d>2$ this quantity is larger than $d^{2}-1$ --  the number of independent parameters required to define physical properties of a Hamiltonian (the term -1 comes from the fact that energy is always defined up to some constant).
	So the properties 2 and 3 are insufficient to determine a set of possible probability representations of Hamiltonians.
	In what follows we consider a necessary and sufficient condition on matrix $\Hbf \in \Mat{d^{2}}{d^{2}}{\mathbb{R}}$ to be a probability representation of some Hamiltonian $H$.
	
	In order to proceed, we first need to introduce the ``vectorised'' notation for operators. 
	If $ A \in  \TrClass{\Hil}$ is an operator, then it can be written in a form
	\begin{equation}
		A = \sum_{n,m=1}^{d} A_{nm} \ket{n}\bra{m}, \quad A_{nm} = \bra{n} A \ket{m}.
	\end{equation}
	Then the bra- and ket-representations of $A$ will be denoted as
	\begin{equation}
		\kett{A} = \sum_{n,m} A_{nm} \ket{n} \otimes \ket{m}, ~~
		\braa{A} = \sum_{n,m} \overline{A_{nm}} \bra{n} \otimes \bra{m}.
	\end{equation}
	These representations may be understood as raising or lowering index.  
	The inner product between such vectors yields $ \braa{A}\kett{B} = \Tr{A^\dagger B} $. 
	We also note that $B=UAV^{\dagger}$ transforms into $\kett{B} = U\otimes \overline{V} \kett{A}$ with $U,V \in  \TrClass{\Hil}$. 
	
	We introduce tensors $\ebf$ and $\Ebf$ in the form
	\begin{equation}
		\ebf =
		\begin{bmatrix} \kett{ e_1 } & \cdots & \kett{ e_{d^2} } \end{bmatrix},
		\quad
		\Ebf =
		\begin{bmatrix} \braa{E_1} \\ \vdots \\ \braa{E_{d^2}} \end{bmatrix}.
	\end{equation}
	Then one has 
	\begin{equation}
		\kett{\rho} = \ebf p,\quad \Ebf \ebf = \mathbf{1}, \quad \ebf \Ebf = \sum_{i=1}^{d^2} \kett{e_i}\braa{E_i}.
	\end{equation} 
	By using this notation, the matrix $\Hbf$ can be written as follows:
	\begin{equation}
		\Hbf = - \imath \Ebf \left( H \otimes \mathbf{I}_{d} - \mathbf{I}_{d} \otimes \overline{H}  \right) \ebf.
	\end{equation}
	
	Let $ \{ \sigma^{(i)} \}_{i=1}^{d^2-1} $ be a linearly independent set of operators, satisfying following conditions:
	(i) operators are traceless  $\Tr{\sigma^{(i)}}=0$;
	(ii) operators are Hermitian $ \sigma^{(i)} = (\sigma^{(i)})^*$;
	(iii)  $ \Tr{\sigma^{(i)} \sigma^{(j)}} = 2 \delta_{ij} $.
	In a two-dimensional Hilbert space ($d=2$) these such a set can be presented with Pauli matrices. 
	Then the Hamiltonian can be represented as follows:
	\begin{equation}
		H = \nu_0 \mathbf{I}_{d} + \sum_{i=1}^{d^2-1} \nu_i \sigma^{(i)},
	\end{equation}
	where $ \nu_0 = \Tr{H}/d$ and $\nu_i = \Tr{H \sigma^{(i)}}/2 $. 
	
	Let $\{\Hbf^{(i)}\}_{i=1}^{d^{2}}$ be a set of matrices corresponding to $ \{ \sigma^{(i)} \}_{i=1}^{d^2-1} $:
	\begin{equation}
		\Hbf^{(i)}_{lk} =  \imath \Tr{\sigma^{(i)} \Commutator{E_l}{e_k}}.
	\end{equation}
	We note the probability representation of a Hamiltonian equal to identity matrix gives zero matrix.
	Therefore, we have
	\begin{equation}
		\Hbf = \sum_{i=1}^{d^2-1} \nu_i \Hbf^{(i)}.
	\end{equation}
	Next, we can see that
	\begin{multline}
	\Tr{ \Hbf^{(i)} \Hbf^{(j)} } \\
	= -\Tr{  \sigma^{(i)} \sigma^{(j)} \otimes \mathbf{I}_{d} + \mathbf{I}_{d} \otimes \overline{\sigma^{(i)}} \overline{\sigma^{(j)}}  } = -4 d \delta_{ij}.
	\end{multline}
	
	Thus, it is possible to define the projector $\mathcal{P}_{\rm unit}(\cdot)$ on the space $\Mat{d^{2}}{d^{2}}{\mathbb{R}}$ that correspond to Hamiltonians in the following explicit form:
	\begin{equation}
		\mathcal{P}_{\rm unit}(\mathbf{M}) = -\frac{1}{4 d} \sum_{i=1}^{d^2-1} \Tr{ \mathbf{M} \cdot \mathbf{H}^{(i)} } \mathbf{H}^{(i)}
	\end{equation}
	(here $\mathbf{M}$ is some $d^{2}\times d^{2}$ matrix ).
	Finally, the matrix $\Hbf$ corresponds to a Hamiltonian, if and only if it satisfies the identity 
	\begin{equation}
		\mathcal{P}_{\rm unit}(\Hbf) = \Hbf.
	\end{equation}
	It also worth to note the $\mathcal{P}_{\rm unit}$ turns to be a useful tool for studying experimental data in quantum process tomography experiments, since it allows extracting the unitary part of generator for a reconstructed process~\cite{Kiktenko2020}.
	
	\subsection{GKSL equation}
	
	Here we generalize the results of the previous section on the case of the GKSL equation~\cite{Gorini1976,Linblad1976}.
	Consider the Markovian master equation in the form $\dot{\rho}(t)=L(\rho)$ with 
	\begin{multline}\label{eq:GKSL}
		L(\rho) = -\imath \Commutator{H}{\rho (t)} + \\
		\sum_k\left(A_k\rho(t) A_k^{\dagger} - \frac{1}{2} \AntiCommutator{A_k^{\dagger}A_k}{\rho(t)} \right),
	\end{multline}
	where  $\AntiCommutator{\cdot}{\cdot}$ is an anticommutator, and $A_k$ are some arbitrary operators describing dissipative evolution, also known as noise operators.
	
	Let us introduce a CP map $\Psi: M \mapsto \sum_{k}A_{k} M A_{k}^{\dagger}$.
	One can think about $\Psi(\cdot)$ as a quantum channel without trace-preserving property.
	The second term of Eq.~\eqref{eq:GKSL} can be written in the following form:
	\begin{equation}
		\Psi(\rho) - \frac{1}{2} \AntiCommutator{\Psi^*(\mathbf{I}_{d})}{\rho} \equiv D(\rho),
	\end{equation}
	where $\Psi^{*}$ is a map dual to $\Psi$.
	
	Let ${\bf S}\in\Mat{d^{2}}{d^{2}}{\mathbb{R}}$ be MIC-POVM-based representation of $\Psi$, i.e. ${\bf S}_{ij}=\Tr{E_{i} \Psi(e_{j})}$.
	Then by using Eqs.~\eqref{eq:channel-from-matr}, \eqref{eq:dual-channel-from-matr}, and \eqref{eq:lambda-tilde} one obtains
	\begin{equation}
 		\Tr{E_{i} D(e_{j})} = {\bf S}_{ij} - \sum_{k,l=1}^{d^{2}}{\bf S}_{kl} \frac{\widetilde\Lambda_{i,l}^{(j)} + \widetilde\Lambda_{l,i}^{(j)}}{2} \equiv {\bf D}_{ij}
	\end{equation}
	for $i,j=1,\ldots,d^{2}$.
	Using the probability representation of the first term of \eqref{eq:GKSL} from the previous section, we obtain the GKSL equation in the MIC-POVM-based probability as follows:
	\begin{equation}
		\dot{p}(t) = \Lbf p(t), \quad \Lbf = \Hbf + \Dbf.
	\end{equation}	
	One can easily verify that $\Dbf \in \Mat{d^{2}}{d^{2}}{\mathbb{R}}$ and $\sum_{i=1}^{d^{2}}\Dbf_{ij}=0$ for every $j=1,\ldots,d^{2}$.
	
	We also discuss a necessary and sufficient properties of ${\bf L}$ to be corresponded to some Liouvillian $L(\cdot)$.
	It is known (see Ref.~\cite{Wolf2012}) that $L(\cdot)$ is a generator of CPTP maps if and only if
	\begin{equation} \label{eq:L-check}
		\overline{P}_{\sigma} ({\rm Id} \otimes L)(\sigma) \overline{P}_{\sigma} \geq 0,
	\end{equation}
	where $\sigma \in \States{\Hil \otimes \Hil}$ is a maximally entangled state [see e.g. Eq.~\eqref{eq:max-ent-state}] and $\overline{P}_{\sigma}=I_{d^{2}}-\sigma$.
	We then consider $d^{4}$-dimensional vectors $s$ and $p_{s}$ with the following components:
	\begin{equation}
		\begin{aligned}
			s_{(ij)} &= \Tr{E_{i}\otimes E_{j} \sigma}, \\ 
			\overline{p}_{s,(ij)} &= \Tr{E_{i}} \Tr{E_{j}} - s_{(ij)}.
		\end{aligned}
	\end{equation}
	The probability representation of the expression in left-hand side of Eq.~\eqref{eq:L-check} takes the form
	\begin{equation}
		 \overline{p}_{s} * ({\bf I}_{d^{2}} \otimes {\bf D}) s * \overline{p}_{s} \equiv p_{\bf D}.
	\end{equation}

	Thus, one can employ the algorithm from Sec.~\ref{sec:states} to check that $p_{\bf D}$ corresponds to non-normalized state with respect to the MIC POVM $E\otimes E$, and thus condition~\eqref{eq:L-check} is fulfilled.
	
	\subsection{Heisenberg picture}
	Up to this point, we described evolution equation for states from the viewpoint of the Schr{\"o}dinger picture.
	Here we show how to adapt the MIC-POVM-based probability representation to the Heisenberg picture, where measurement operators evolve.
	As a master equation, we consider GKSL equation from the previous section.

	Consider a POVM $M=\{M_{i}\}_{i=1}^{m}$.
	Remember, that in the MIC-POVM-based representation it is defined by $m \times d^{2}$ pseudostocastic matrix ${\bf M}$.
	The probability vector of measurement outcomes at time $t\geq 0$ is given by
	\begin{equation}
		{\bf M}p(t) = {\bf M}e^{{\bf L}t}p_{0}.
	\end{equation}
	Now we can introduce a Heisenberg representation of ${\bf M}$:  ${\bf M}^{\rm Heis}(t):={\bf M}e^{{\bf L}t}$, that is solution of the equation
	\begin{equation}\label{eq:Heis-GKSL}
		\frac{d}{dt}{\bf M}^{\rm Heis}(t) = {\bf M}^{\rm Heis}(t){\bf L}, \quad {\bf M}^{\rm Heis}(0) = {\bf M}(0).
	\end{equation}
	The probabilities of outcomes are given by $q(t) = {\bf M}^{\rm Heis}(t)p_{0}$
	We note that Eq.~\eqref{eq:Heis-GKSL} can be also adapted to an evolution of a particular effect (row of ${\bf M}$) instead of full matrix ${\bf M}$.
	In the case of Hermitian observable one can write the following equation for an row-vector allowing to compute mean value of $\langle O(t)\rangle = {\bf O}_{\rm mean}^{\rm Heis}(t)p_{0}$:
	\begin{equation}\label{eq:Heis-GKSL}
		\frac{d}{dt}{{\bf O}}_{\rm mean}^{\rm Heis}(t) = {\bf O}_{\rm mean}^{\rm Heis}(t) {\bf L}, \quad {\bf O}_{\rm Heis}^{\rm mean}(0) = {\bf O}_{\rm mean}.
	\end{equation}
	
	\section{MIC-POVM probability representation and quantum-to-classical transition}\label{sec:decoherence}
	
	One of the important questions that can be addressed in the probability representation of quantum dynamics via pseudostochastic maps is how to quantify an aspect of `non-classicality' of a particular quantum dynamics.
	As we see, quantum dynamics is essentially different from classical stochastic dynamics by the possibility of negative conditional probabilities.
	The study of negative elements in pseudostochastic matrices seems to be very important in particular for understanding the origin of the complexity of simulating quantum dynamics of large-scale quantum systems. 
	We note that the fact of the complexity of efficient simulation of the behaviour of quantum systems by classical stochastic systems is indeed a widely believed but unproven conjecture.  

	At the same time, the presence of decoherence drastically changes the nature of the dynamics of quantum-mechanical systems. 
	We note that in Ref.~\cite{Kiktenko2020} it has been shown that decoherence process accompanying a unitary quantum dynamics in the framework of quantum Markovian master equation can eliminate negative elements 
	in the resulting pseudostochastic matrix making it purely stochastic and looking like a classical stochastic process.
	Here we study a decay quantum features of a dissipative quantum Markovian dynamics within a MIC-POVM representation and compare the cases of SIC-POVM-based and general MIC-POVM-based representations.
	
	First of all, we note that appearance of negative conditional probabilities in pseudostochastic matrix ${\bf S}(t)=e^{{\bf L}t}$ is determined by negative non-diagonal elements of the generator ${\bf L}$.
	On the one hand, if there exist at least one negative non-diagonal element ${\bf L}_{i^{\star}j^{\star}}<0$ ($i^{\star}\neq j^{\star})$, 
	then there appear negative elements in ${\bf S}(t)$ at least for small enough time $t>0$ for which ${\bf S}(t)\approx {\bf I}+{\bf L}t$.
	On the other hand, if all non-diagonal elements of ${\bf L}$ are non-negative, then considering the identity 
	\begin{equation}
		{\bf S}(t) = \lim_{n\rightarrow\infty}  \left( {\bf I}+{\bf L}\frac{t}{n} \right)^{n}
	\end{equation}
	we come to the conclusion that all elements of ${\bf S}(t)$ are non-negative for any $t>0$ since all elements of ${\bf I}+{\bf L}t/n$ are non-negative.

	Let us have in mind that  (i) for every generator of the GKSL equation one has $\sum_{i}{\bf L}_{ij}=0$, and (ii) in the case of purely unitary evolution one has the diagonal elements being equal to zero.
	From these two facts it directly follows that every non-trivial generator of decoherence-free evolution ${\bf L}={\bf H}\neq 0$ has at least one negative element for sure.
	As it was already mentioned, an appearance of a dissipator term ${\bf D}$ in the generator ${\bf L}={\bf H}+{\bf D}$ can change the situation, and negative elements can disappear from the generator.
	
	We then consider the case of a spin-1/2 particle processing in a magnetic field and exposed by one the following standard dissipative process: depolarizing, dephasing and amplitude damping.
	From the viewpoint of the standard formalism, we consider a Hamiltonian in the following form:
	\begin{equation}
		H_{\theta} = \frac{1}{2}(\sigma^{(1)}\sin\theta + \sigma^{(3)}\cos\theta),
	\end{equation}
	and the following variants of noise-operators (Lindblad operators):
	\begin{eqnarray}
		&A_{\tau,i}^{\rm depol}&= \frac{1}{2\sqrt{\tau}}\sigma^{(i)}, \quad i=1,2,3;\\
		&A_{\tau,0}^{\rm deph}&= \frac{1}{\sqrt{\tau}}\sigma^{(3)},\\
		&A_{\tau,0}^{\rm damp}&= \frac{1}{\sqrt{\tau}}\sigma^{-}, \quad \sigma^{-}=\frac{1}{2}(\sigma^{(1)}-\imath\sigma^{(2)}).
	\end{eqnarray}
	Here $\sigma^{(1)}, \sigma^{(2)}, \sigma^{(3)}$ are standard $x$-, $y$-, $z$- Pauli matrices correspondingly, $\theta$ is a real polar angle which determines a direction of the magnetic field (azimuthal angle is set to zero), and $\tau>0$ is a decoherence process characteristic time.
	Thus we consider the GKSL equation $\dot{\rho}= L_{\theta,\tau}^{\rm dec}[\rho]$ with generator
	\begin{multline}\label{eq:mainGenerator}
		L_{\theta,\tau}^{\rm dec}[\rho]=-\imath \Commutator{H_{\theta}}{\rho} + \\\sum_k\left(A_{\tau,k}^{\rm dec}\rho A_{\tau,k}^{\rm dec\dagger} - \frac{1}{2} \left[A_{\tau,k}^{\rm dec\dagger}A_{\tau,k}^{\rm dec}\rho + \rho A_{\tau,k}^{\rm dec\dagger}A_{\tau,k}\right]\right)
	\end{multline}
	with ${\rm dec}\in\{ {\rm depol}, {\rm deph}, {\rm damp} \}$.
	We note that Larmor frequency is fixed and equal to 1 and all three decoherence processes are invariant under rotation around $z$-axis.
	
	Next, we study a pseudostochastic matrix of the quantum channel, which maps an initial state at the moment $t=0$ to the final state at some $t>0$ according to Eq.~\eqref{eq:mainGenerator}.
	In particular, we are interested in the conditions on the parameters $\theta$ and $\tau$ which make the resulting evolution matrix to be purely stochastic for any time $t>0$.
	As it was already mentioned, this kind of quantum-to-classical transition is determined by the form of the generator.
	In turn, it depends not only on the values of $\theta$ and $\tau$, but also on the particular MIC-POVM used for its representation.

	\begin{figure*}
		\includegraphics[width=\linewidth]{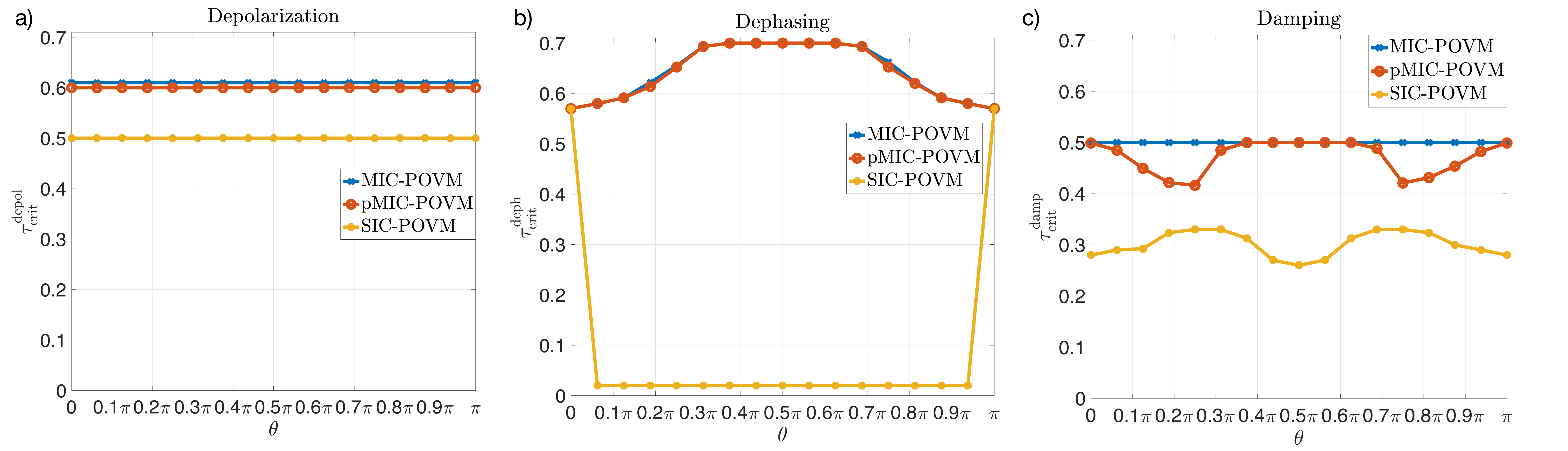}
		\caption{Results of numerical calculations of the  characteristic time $\tau_{\rm crit}^{\rm dec}$, such that the resulting evolution looks ``classical-like'', for three different decoherence channels and different sets of MIC-POVM: in (a) depolarization, in (b) dephasing, and in (c) damping channels.
		One can see that the MIC-POVM-based probability representation gives more strict requirements for revealing the non-classical character of dissipative quantum dynamics in compare with the SIC-POVM-based approach.}
		\label{fig:tau_crit}
	\end{figure*}

	Let ${\bf L}^{[E]}$ be a probability representation of generator~\eqref{eq:mainGenerator} with respect to a MIC-POVM $E$.
	As a reference point, we take a SIC-POVM $E^{\rm sym}$ with the following effects:
	\begin{equation}\label{eq:SIC-POVM}
		\begin{aligned}
			E_1^{\rm sym} &= \frac{1}{4} {\bf I}_{2} + \frac{\sqrt{3}}{12} (-\sigma^{(1)} + \sigma^{(2)} + \sigma^{(3)}), \\
			E_2^{\rm sym} &= \frac{1}{4} {\bf I}_{2} + \frac{\sqrt{3}}{12} (\sigma^{(1)} - \sigma^{(2)} + \sigma^{(3)}), \\
			E_3^{\rm sym} &= \frac{1}{4} {\bf I}_{2} + \frac{\sqrt{3}}{12} (\sigma^{(1)} + \sigma^{(2)} - \sigma^{(3)}), \\
			E_4^{\rm sym} &= \frac{1}{4} {\bf I}_{2} + \frac{\sqrt{3}}{12} (- \sigma^{(1)} - \sigma^{(2)}- \sigma^{(3)}).
		\end{aligned}
	\end{equation}	
	In the ${\bf E}^{\rm sym}$-based representation the generator term corresponding to the unitary evolution takes the form
	\begin{multline}
		{\bf H}_{\theta}^{[E^{\rm sym} ]} = \frac{\sin\theta}{4}
			\begin{bmatrix}
				0  & -1 & 1 & 0 \\ 
				1  &  0 & 0 & -1 \\
				-1 & 0 & 0  & 1 \\
				0  & 1 & -1 & 0 \\
			\end{bmatrix}
		+ \\\frac{\cos\theta}{4}
			\begin{bmatrix}
				0  & -1 & 0 & 1 \\ 
				1  &  0 & -1 & 0 \\
				0 & 1 & 0  & -1 \\
				-1  & 0 & 1 & 0 \\
			\end{bmatrix}.
	\end{multline}
	The generators of depolarization, dephazing and damping are
	\begin{eqnarray}
	&{\bf D}^{{\rm depol} [E^{\rm sym}]}_{\tau} &= \frac{1}{4\tau}
			\begin{bmatrix}
				-3  & 1 & 1 & 1 \\ 
				1 &  -3 & 1 & 1 \\
				1 & 1 & -3  & 1 \\
				1 & 1 & 1 & -3 \\
			\end{bmatrix},\\
	&{\bf D}^{{\rm deph}[E^{\rm sym}]}_{\tau} &= \frac{1}{2\tau}
			\begin{bmatrix}
				-1  & 1 & 0 & 0 \\ 
				1 &  -1 & 0 & 0 \\
				0 & 0 & -1  & 1 \\
				0 & 0 & 1 & -1 \\
			\end{bmatrix},\\
	&{\bf D}^{{\rm damp}[E^{\rm sym}]}_{\tau} &=
			\frac{1}{4\tau}
			\begin{bmatrix}
				-2  & 0 & 1 & 1 \\ 
				0 &  -2 & 1 & 1 \\ 
				1& 1 & -2 & 0 \\
				1 & 1 & 0 & -2 \\
			\end{bmatrix} + \\&&\frac{1}{4\sqrt{3}\tau}
			\begin{bmatrix}
				1 & 1 & 1 & 1 \\ 
				1 &  1 & 1 & 1 \\ 
				-1& -1 & -1 & -1 \\
				-1 & -1 & -1 & -1 \\
			\end{bmatrix},
	\end{eqnarray}	
	respectively.
	
	The transition of the generators to an arbitrary MIC-POVM $E$-based representation can be realized via the transformation
	\begin{equation}
		{\bf L}_{\theta,\tau}^{{\rm dec}[E]} = {\bf M}_E^{[E_{\rm sym}]} {\bf L}_{\theta,\tau}^{{\rm dec}[{E^{\rm sym} }]} {\bf M}_{E^{\rm sym}}^{[E]},
	\end{equation}
	where ${\bf L}_{\theta,\tau}^{\rm dec[\cdot]}={\bf H}_{\theta}^{[\cdot]}+{\bf D}_{\tau}^{\rm dec[\cdot]}$ and ${\bf M}^{[E_{2}]}_{E_{1}}$ is a pseudostochastic matrix of a measurement with MIC-POVM $E_{1}$ in MIC-POVM $E_{2}$-based representation.
	
	Let  $\mathcal{N}({\bf L}^{[E]})$ be a sum of negative elements in ${\bf L}^{[E]}$:
	\begin{equation}
		\mathcal{N}({\bf L}^{[E]}) = \sum_{i,j} \max \left(0, -{\bf L}^{[E]}_{ij}\right), \quad  i \neq j.
	\end{equation}	
	This quantity characterizes an extent of non-classicality of a quantum Markovian generator with respect to the probability representation with MIC-POVM  $E$.
	The crucial point is that $\mathcal{N}({\bf L}^{[E]})$ in the case of the considered spin-1/2 evolution can depend not only on physical parameters $\tau$ and $\theta$, but also on the employed MIC-POVM $E$.
	In order to reduce dependence on particular MIC-POVM for quantifying non-classicality, we introduce the following quantity:
	\begin{equation}
		\mathcal{N}_{\Omega}(L) = \min_{E\in\Omega}\mathcal{N}({\bf L}^{[E]}),
	\end{equation}
	where $\Omega$ is the set of MIC-POVMs, $L$ is the generator of the quantum Markovian dynamics (one can think of it as a super-operator from the right-hand side of the GKSL equation written in the standard form), 
	and ${\bf L}^{[E]}$ is the $E$-based probability representation of $L$.
	Finally, we introduce a quantity
	\begin{equation}
		\tau_{\rm crit}^{\rm dec}(\theta; \Omega) = \sup\Set{\tau}{\mathcal{N}_{\Omega}(L_{\theta,\tau}^{\rm dec})=0},
	\end{equation}
	which shows a minimal strength of decoherence process ${\rm dec}\in\{{\rm depol}, {\rm deph}, {\rm damp}\}$, which is characterized by corresponding characteristic time, such that the resulting evolution looks ``classical-like'' 
	(i.e. has stochastic evolution matrix) from the viewpoint of optimal probability representation with MIC-POVM taken from some set $\Omega$.
	We note that this value also depends on the angle $\theta$ which determines a unitary evolution process.
	
	We calculate $\tau_{\rm crit}^{\rm dec}(\theta; \Omega)$ numerically for three different sets $\Omega$: 
	(i) the set of all possible SIC-POVM;
	(ii) the set of projective MIC-POVMs (pMIC-POVMs), that is MIC-POVMs consisted of rank-1 projectors only; and 
	(iii) the general set of all possible MIC-POVMs.
	One can see that these three cases go from a special one to the most general one.
	The results of our numerical calculations are presented in Fig.~\ref{fig:tau_crit}.
	
	As one may expect, for the depolarization model [see Fig.~\ref{fig:tau_crit}(a)] a dependence on $\theta$ is absent due to the symmetry of the corresponding generator.
	The obtained values of $\tau_{\rm crit}^{\rm depol}$ for SIC-POVM, pMIC-POVM, and MIC-POVM cases are 0.5, 0.6, and 0.61 correspondingly.
	So the transition from SIC-POVM-based representation to MIC-POVM-based representation allows decreasing a tolerable strength of depolarization process at which the whole quantum process looks similar to a classical stochastic process.
	Dependence on the angle $\theta$ appears in the dephasing process. 
	Fig.~\ref{fig:tau_crit}(b) shows that MIC-POVMs provide the same results as MIC-POVMs, while SIC-POVMs allows obtaining the stochastic form of the evolution matrix only in the small neighbourhood of $\theta=0$ and $\theta=\pi$.
	In the case of damping [see Fig.~\ref{fig:tau_crit}(c)] we see that pMIC-POVMs and MIC-POVMs provide results better than SIC-POVMs, 
	meanwhile we obtain $\tau_{\rm crit}^{\rm damp} = 0.5$ for any $\theta$ whereas results for pMIC-POVMs depend on $\theta$ with some plateau-like behavior around $\theta = 0.5$.
	
	On the basis of the obtained results, we can conclude that turning from the SIC-POVM representation to the MIC-POVM representation allows pushing back a  `quantum-classical border' from the side of classical processes, 
	and thus makes it easier to employ a theory of classical stochastic processes to quantum dynamics.
	These results are relevant to future investigation of emulating quantum dynamics with randomized algorithms run on a classical computer and, correspondingly, to the quantum advantage problem~\cite{Martinis2019,Wisnieff2020,Chen2020,Lidar2020}.
	
	\section{MIC-POVM-based representation for quantum computing}\label{sec:qcomp}
	
	In this section, we discuss how the MIC-POVM representation allows taking a fresh look at the process of quantum computing within a (digital) gate-based quantum computing model with qubits.
	In this model, execution of quantum algorithm can be divided into three basic steps: (i) initialization of a qubit register, (ii) manipulation with qubit states, and (iii) read-out measurement.
	The state initialization essentially represents the preparation of a state $\ket{\psi^{\rm init}}=\ket{0}^{\otimes n}$, 
	where $n$ is a number of qubits (hereinafter we use standard notation $\{\ket{0}, \ket{1}\}$ for the computational basis vectors of Hilbert spaces corresponding to each qubit).
	The state manipulation is represented as a sequence of single-qubit and two-qubit gates, i.e. a set of unitary operators acting in corresponding Hilbert spaces.
	
	It is a well-known fact that any quantum algorithm can be efficiently decomposed into a sequence of gates from some finite universal gate set (see Ref.~\cite{NielsenChuang}), 
	which consists of a single two-qubit gate, such as controlled-NOT (CNOT) or controlled-phase gates:
	\begin{equation}
		CX = \begin{bmatrix}
			1 & 0 & 0 & 0 \\
			0 & 1 & 0 & 0 \\
			0 & 0 & 0 & 1 \\
			0 & 0 & 1 & 0 \\
		\end{bmatrix}, \quad
		CZ = \begin{bmatrix}
			1 & 0 & 0 & 0 \\
			0 & 1 & 0 & 0 \\
			0 & 0 & 1 & 0 \\
			0 & 0 & 0 & -1 \\
		\end{bmatrix}, \quad
	\end{equation}
	and number of single-qubit gates, e.g. Hadamard gate $H$ and $T$-gate:
	\begin{equation}
		H=\frac{1}{\sqrt{2}}\begin{bmatrix}
			1 & 1\\
			1 & -1
		\end{bmatrix}, \quad
		T=\begin{bmatrix}
			1 & 0 \\
			0 & e^{\imath  \pi/4}
		\end{bmatrix}.
	\end{equation}
	The read-out measurement is essentially a projective measurement in the computation basis performed on some subset (or the full set) of $n$ qubits.
	One can think about the read-out measurement as a sampling of a random variable according to a distribution determined by the state $\ket{\psi^{\rm fin}}=U^{\rm circ}\ket{\psi^{{\rm init}}}$, where $U^{\rm circ}$ is unitary operator of all applied gates.
	
	\subsection{Initialization}
	
	\begin{figure}
		\includegraphics[width=0.9\linewidth]{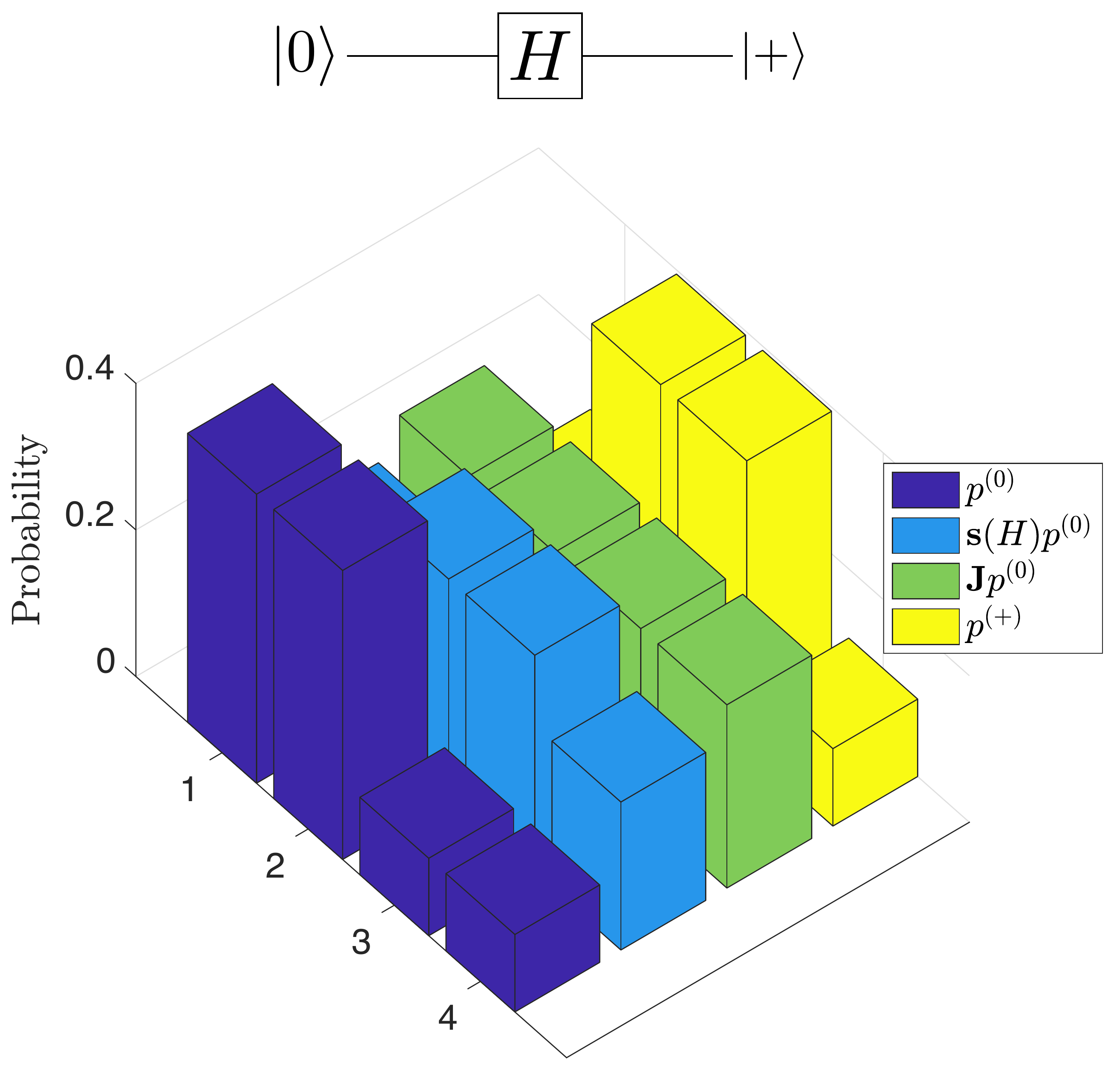}
		\vskip -6mm
		\caption{Evolution of the single qubit state $\ket{0}$, which is affected by the Hadamard gate, in the probability representation. Label $i\in\{1,2,3,4\}$ corresponds to the effect $E_{i}^{\rm sym}$.}
		\label{fig:had_gate_impl}
	\end{figure}
	
	\begin{figure*}
		\includegraphics[width=1.0\linewidth]{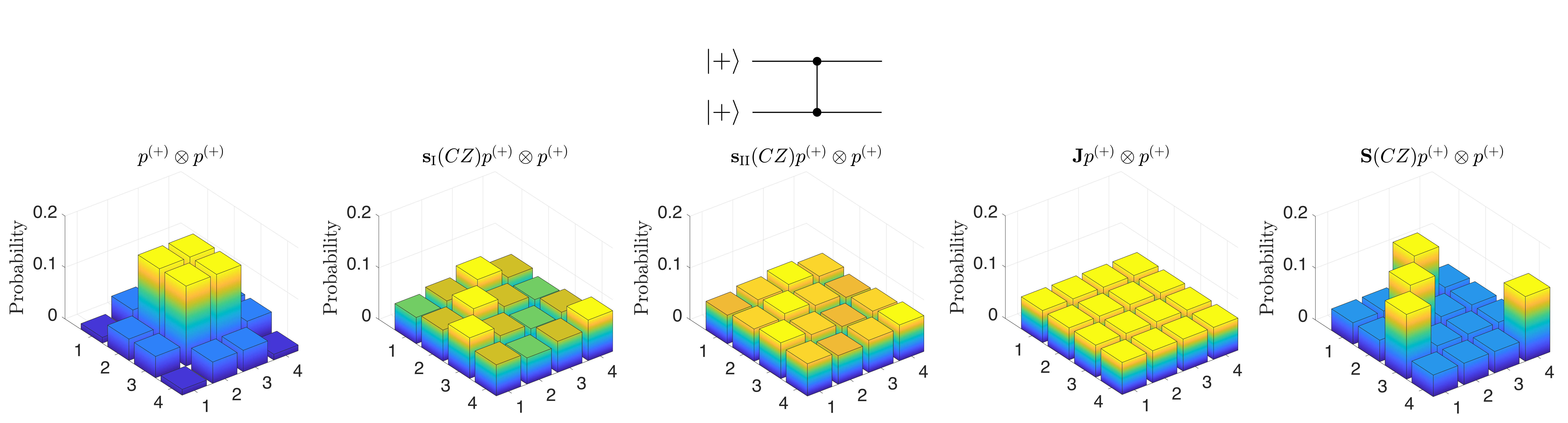}
		\caption{Construction of the cluster state in the probability representation. Here $(ij)$ outcome corresponds to $E_{i}^{\rm sym}\otimes E_{j}^{\rm sym}$ effect of the employed MIC-POVM~\eqref{eq:symmetric-MIC-POVM}.}
		\label{fig:cz_gate_impl}
	\end{figure*}
	
	Let us consider a process of quantum computations from the viewpoint of MIC-POVM based probability representation.
	Within this section we fix a single MIC-POVM $E$ constructed with tensor product operation from SIC-POVM effects~\eqref{eq:SIC-POVM}:
	\begin{equation}\label{eq:symmetric-MIC-POVM}
		E = \{E_{i_{1}}^{\rm sym} \otimes \ldots \otimes E_{i_{n}}^{\rm sym} \}_{i_{1},\ldots,i_{n}\in\{1,2,3,4\}}.
	\end{equation}
	The initial  $4^{n}$-dimensional probability vector corresponding to $\ket{\psi^{\rm init}}$ then takes the form
	\begin{equation}
		p^{\rm init} = p^{(0)}\otimes \ldots \otimes p^{(0)},
	\end{equation}
	where 
	\begin{equation}
		p^{(0)}=\begin{bmatrix} \frac{3+\sqrt{3}}{12} & \frac{3+\sqrt{3}}{12} & \frac{3-\sqrt{3}}{12} & \frac{3-\sqrt{3}}{12}   \end{bmatrix}^\top
	\end{equation}
	is probability vector corresponding to the state $\ket{0}$.
	We note that in the considered probability representation each qubit corresponds to a classical two-bit string, and the whole $n$-qubit state can be considered as a probability distribution over all possible values of $2n$-bit string.
	
	\subsection{Single-qubit gates}
	We consider a representation of single-qubit and two-qubit gates.
	A pseudostochastic matrix of a single-qubit gate $U_{1}$ in the SIC-POVM representation reads:
	\begin{equation}\label{eq:singlequbit}
		{\bf S}({U_{1}})=3{\bf s}({U_{1}})-2{\bf J}_{1},
	\end{equation}
	where
	\begin{equation}
	\begin{split}
		{\bf s}({U_{1}})_{ij} = 2{\rm Tr}(E^{\rm sym}_{i}U_{1}E^{\rm sym}_{j}U_{1}^\dagger), \\ \quad i,j=1,2,3,4,
	\end{split}
	\end{equation}
	and ${\bf J}_{1}$ is a $4\times 4$ matrix with all entities equal to 1/4.
	It is easy to see that ${\bf s}({U_{1}})$  is a bistochastic matrix (note that $\rho_{j}$ can be considered as a fair quantum state).
	The matrix ${\bf J}_{1}$ is also bistochastic and outputs maximally chaotic state for any input probability vector.
	Pseudostochastic matrices for some common single-qubit gates are provided in Appendix~\ref{apx:gates}.
	We also demonstrate an application of the Hadamard gate to the state $\ket{0}$ in Fig.~\ref{fig:had_gate_impl}.
	
	\subsection{Multi-qubit gates}

	In the case of $n\geq 2$ qubits, the pseudostochastic matrix corresponding to an action of $U_{1}$ on a particular qubit can be obtained by tensor product with identity matrix (matrices).
	As an illustrative example consider the case where $U_{1}$ acts on the second qubit among $n=4$ qubits.
	Since an absence of operation corresponds to identity stochastic matrix the resulting pseudostochastic matrix reads
	\begin{multline}\label{eq:exampleofmultiplequbits}
		{\bf I}_{4} \otimes {\bf S}({U_{1}}) \otimes {\bf I}_{4}\otimes {\bf I}_{4}\\=3 {\bf I}_{4} \otimes  {\bf s}({U_{1}})\otimes {\bf I}_{16}-2{\bf I}_{4} \otimes {\bf J}_{1}\otimes {\bf I}_{16}.
	\end{multline}
	We note that the right-hand side of Eq.~\eqref{eq:exampleofmultiplequbits} is a difference of two scaled stochastic matrices.
	
	We see that the probability vector resulting from the action of the considered pseudostochastic matrix appears to be a linear combination of two probability vectors obtained from the initial one by acting with different bistochastic matrices.
	The sum of coefficients of this linear combination is equal to one thus forming affine combinations.
	Due to the negative elements inside, it is very different from a convex hull typical for a classical randomization process.
	
	The situation with a two-qubit entangling gate $U_{2}$ (e.g. $ CX$ or $CZ$) is a bit more complex.
	The corresponding pseudostochastic matrix reads
	\begin{equation}\label{eq:twoqubit}
		{\bf S}(U_{2})= 9{\bf s}_{\rm I}(U_{2}) - 12 {\bf s}_{\rm II}(U_{2}) + 4 {\bf J}_{2}, 
	\end{equation}
	where 
	\begin{equation}
		\begin{aligned}
			&{\bf s}_{\rm I}(U_{2})_{(ij),(kl)}=4{\rm Tr}(E^{\rm sym}_{i}\otimes E^{\rm sym}_{j} U_{2}E^{\rm sym}_{k}\otimes E^{\rm sym}_{l}U_{2}^{\dagger}),	\\
			&{\bf s}_{\rm II}(U_{2})_{(ij),(kl)}={\rm Tr}(E^{\rm sym}_{i}\otimes E^{\rm sym}_{j} U_{2}\widetilde{\rho}_{kl}U_{2}^{\dagger}),	\\
			&\widetilde{\rho}_{kl}= E^{\rm sym}_{k}\otimes\rho^{\rm mix} + \rho^{\rm mix} \otimes E^{\rm sym}_{l}, \\ 
			&\quad i,j,k,l=1,2,3,4,
		\end{aligned}
	\end{equation}
	where $\rho^{\rm mix}={\bf I}_{2}/2$ is a maximally mixed state, ${\bf J}_{2}$ is a matrix with all entities equal to 1/16.
	One can check that all matrices  ${\bf s}_{\rm I}(U_{2})$, ${\bf s}_{\rm II}(U_{2})$, and ${\bf J}_{2}$ are bistochastic.
	We observe again a linear combination of stochastic matrices with coefficient giving a total of unity and having negative elements.
	As an example we show the construction of a two-node cluster state in Fig.~\ref{fig:cz_gate_impl}.
	We also provide pseudostochastic matrices corresponding to different two-qubit gates in  Appendix~\ref{apx:gates}.
	The pseudostochastic matrix of two-qubit operation acting on a particular pair of $n\geq 3$ qubits can be obtained by employing tensor product with identity matrix (matrices) in a similar way as in the case of the single-qubit gate.

	\subsection{Measurements}
	
\begin{figure}
		\includegraphics[width=0.7\linewidth]{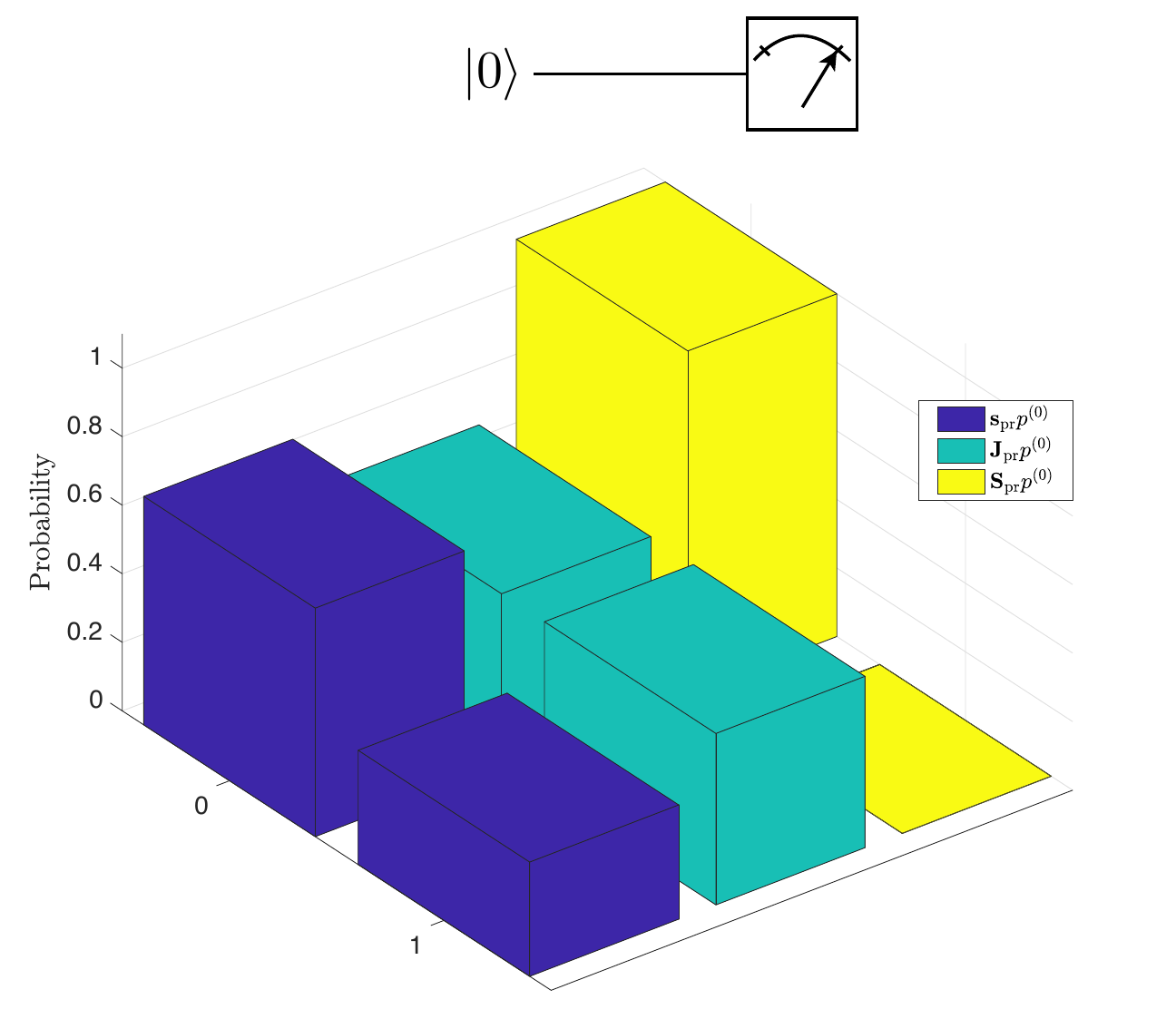}
		\caption{Projective measurement in the probability representation. Here 0 and 1 stand for standard outcome labels.}
		\label{fig:proj_meas}
	\end{figure}

	The pseudostochastic matrix of the single-qubit projective measurement in the computation basis is given by the following expression:
	\begin{equation}\label{eq:projmeas}
		{\bf M}_{\rm pr} = 3{\bf m}_{\rm pr} - 2{\bf J}_{\rm pr},
	\end{equation}
	where
	\begin{equation}
		{\bf m}_{\rm pr} = \frac{1}{2}
			\begin{bmatrix}
				1+\frac{1}{\sqrt{3}} & 1+\frac{1}{\sqrt{3}}  & 1-\frac{1}{\sqrt{3}} & 1-\frac{1}{\sqrt{3}} \\
				1-\frac{1}{\sqrt{3}} & 1-\frac{1}{\sqrt{3}}  & 1+\frac{1}{\sqrt{3}} & 1+\frac{1}{\sqrt{3}} 
			\end{bmatrix}
	\end{equation}
	and ${\bf J}_{\rm pr}$ is $2\times 4$ stochastic matrix with all elements equal to 1/2.
	One can see that the form of Eq.~\eqref{eq:projmeas} is similar to Eq.\eqref{eq:singlequbit}. 
	However, the important difference is that within the projective measurement we have a reduction of probability space dimensionality.
	An example of the projective measurement of a state $\ket{1}$ in the probability representation is shown in Fig.~\ref{fig:proj_meas}.
	In the case of $n>1$ qubits  the pseudostochastic matrix of the single-qubit measurement can be obtained in a similar fashion as in the case single-qubit gate [see example in Eq.~\eqref{eq:exampleofmultiplequbits}]
	A pseudostochasitc matrix of the projective measurement of several qubits can be obtained as a sequence of measurements on individual qubits. 
	
	We see that the running of $n$-qubit quantum circuit can be considered as kind of random walk of $2n$-bit string.
	In each step corresponding to an implementation of single-qubit or two-qubit quantum gates a single or a pair of 2-bit chunks are affected (each chunk consists of $(2k)$th and $(2k+1)$th position in the string with $k=0,1,\ldots,n-1$).
	The final step of readout measurement corresponds to a compressive random mapping of each 2-bit chunk, corresponding to measured qubit, to 1 bit value.
	The quantum nature of this randomized process manifests itself by the fact that all steps are described with pseudostochastic matrices given by Eqs.~\eqref{eq:singlequbit}, \eqref{eq:twoqubit}, and \eqref{eq:projmeas}.
	On the one hand the negative conditional probabilities of pseudostochastic matrices prevent us from straightforward emulating of quantum processes within quantum computation with classical randomized algorithms, 
	and on the hand they actually underlie the advantage of quantum computers over classic ones.
	
	\begin{figure*}
		\centering
		\includegraphics[width=1.0\linewidth]{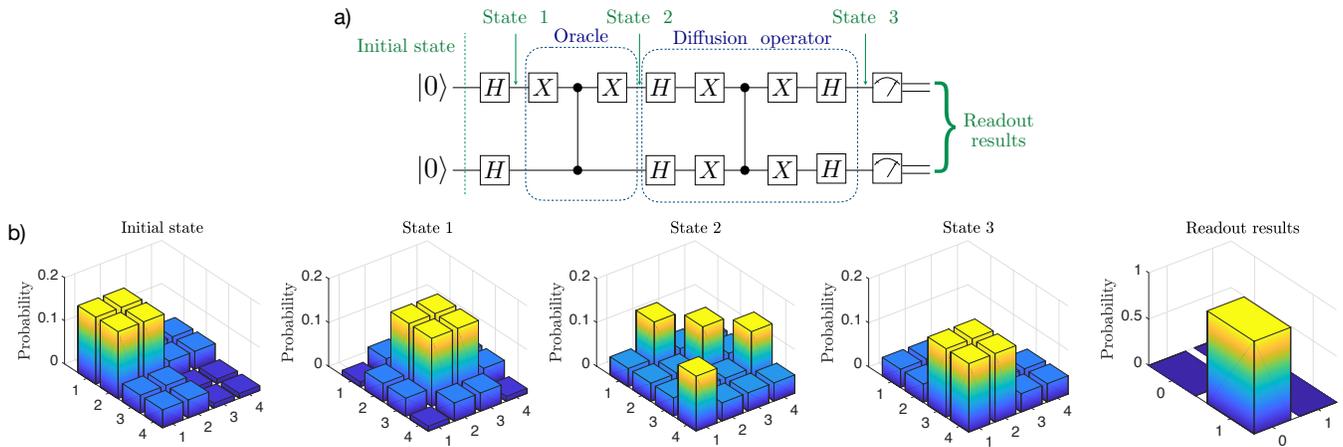}
		\caption{Demonstration of Grover's algorithm in the probability representation. 
		In (a) the circuit for Grover's algorithm is illustrated. 
		In (b) the step-by-step implementation of Grover's algorithm in the probability representation is shown.}
		\label{fig:Grover}
	\end{figure*}
	
	\subsection{Grover's algorithm}
	
	In order to provide another illustrative example of the probabilistic representation of quantum computing processes, we consider a two-qubit Grover's algorithm~\cite{Grover1996} with a classical oracle function 
	\begin{equation}
		 \chi(10)=1, \quad \chi(00) = \chi(01) = \chi(11) = 0.
	\end{equation}
	The corresponding circuit which allows finding the `secret string' 10 by a single query to the two-qubit quantum oracle 
	\begin{equation}
		U_{\chi}\ket{xy}=(-1)^{\chi(xy)}\ket{xy}, \quad x,y\in\{0,1\}
	\end{equation}
	is shown in Fig.~\ref{fig:Grover}(a).
	The evolution of the probability representation of corresponding four-bit string is shown in Fig.~\ref{fig:Grover}(b).
	One can observe how the information about the secret string first gets into the state after applying the oracle and then is extracted with the diffusion gate and projective measurement.

	\section{Conclusion and outlook}\label{sec:conclusion}

	Here we summarize the main results of our paper.
	We have developed the MIC-POVM-based probability representation, which generalizes the SIC-POVM-based approach. 
	We have demonstrated advantages of this approach with the focus on the description of multi-qubit systems.
	We have derived  quantum dynamical equations both for the unitary von-Neumann evolution and the Markovian dissipative evolution, which is governed by the Gorini-Kossakowski-Sudarshan-Lindblad (GKSL) generator.
	We have also discussed applications of the suggested approach for the analysis of NISQ computing processes and obtain pseudostochastic maps for various decoherence channels and quantum gates.
	In particular, we have demonstrated that the MIC-POVM-based probability representation gives more strict requirements for revealing the non-classical character of dissipative quantum dynamics in comparison with the SIC-POVM-based approach. 
	These results seem to be relevant to future investigation of emulating quantum dynamics with randomized algorithms run on a classical computer and, correspondingly, 
	to the quantum advantage problem~\cite{Martinis2019,Wisnieff2020,Chen2020,Lidar2020}.
	Our approach seems to be promising in the context of investigating dynamics of open quantum systems with initially correlated states of a system and bath, which has been studied in details in Ref.~\cite{Wiseman2019}.
	
	We note that there is an interesting connection between MIC representations (and SIC representations particularly) with Lie algebras and Lie groups~\cite{Fuchs2011,Fuchs2015,Manko2010}. 
	The properties of a concrete MIC-POVM may be studied using the structure constants of a generated Lie algebra. For example, it was shown, that the structure constants are completely antisymmetric exactly in the case of SIC-POVMs \cite{Fuchs2015}.

	We also would like to note that the representation may be considered as a faithful functor from the category of channels to the category of pseudostochastic maps (see Ref.~\cite{Wetering2017}), which can be explored in future in more details. 
	Another interesting direction for the further research is related to tensor networks~\cite{Carrasquilla2019, Luchnikov2019}, which are also important from the perspective of quantum computing.

	\smallskip
	
	\section*{Acknowledgments}
	
	We thank D. Chru{\'s}ci{\'n}ski,  A.E. Teretenkov, and V.I. Man'ko for useful comments.
	Results of Secs.~\ref{sec:construction},~\ref{sec:dynamics}, and ~\ref{sec:qcomp} were obtained by E.O.K. and V.I.Y with the support from the Russian Science Foundation Grant No. 19-71-10091.
	Results of Sec.~\ref{sec:decoherence} were obtained by A.K.F. and A.S.M. with the support from the Grant of the President of the Russian Federation (Project No. MK923.2019.2).
	
	Corresponding authors: E.O.K (e.kiktenko@rqc.ru) and A.K.F (akf@rqc.ru).
	
	\smallskip
	\appendix
	
	\section{Pseudostochastic matrices of common single-qubits channels and single- and two-qubit gates}\label{apx:gates}
	We provide explicit form of pseudostochastic matrices for some common types of quantum channels in the case spin 1/2 particle (qubit) in Table~\ref{tbl:generators-matrices}.
	For constructing probability representation of all channels, SIC-POVM ~\eqref{eq:SIC-POVM} is used.
	
	\begin{table*}[h!]
	\caption{Pseudostochastic matrices corresponding to some common channels in SIC-POVM-based formalism. }
	\begin{tabular}{|p{0.2\linewidth}|p{0.28\linewidth}|p{0.45\linewidth}|}
	\hline
	 Quantum channel & Standard formalism & Pseudostochastic matrix \\	\hline
	Identity
	& $ \Phi_t[\rho] = \rho $
	& $
	\begin{bmatrix}
	1 & 0 & 0 & 0 \\
	0 & 1 & 0 & 0 \\
	0 & 0 & 1 & 0 \\
	0 & 0 & 0 & 1
	\end{bmatrix} $\\
	\hline
	Depolarization
	& $ \Phi_t[\rho] = e^{-t/\tau} \rho + (1-e^{-t/\tau}) \chi $
	& $ \frac{1}{4}
	\begin{bmatrix}
	1 + 3 e^{-t/\tau} & 1 - e^{-t/\tau}& 1 - e^{-t/\tau} & 1 - e^{-t/\tau} \\
		1 - e^{-t/\tau} & 1 +3e^{-t/\tau}& 1 - e^{-t/\tau} & 1 - e^{-t/\tau} \\
  	1 - e^{-t/\tau} & 1 - e^{-t/\tau}& 1 +3e^{-t/\tau} & 1 - e^{-t/\tau} \\
  	1 - e^{-t/\tau} & 1 - e^{-t/\tau}& 1 - e^{-t/\tau} & 1 +3e^{-t/\tau}
	\end{bmatrix} $
	\\
	\hline
	Dephasing
	& $ \Phi_t[\rho] =
	\begin{bmatrix}
	\rho_{00} & e^{-t/\tau} \rho_{01} \\
	e^{-t/\tau} \rho_{10} & \rho_{11}
	\end{bmatrix} $
	& $\frac{1}{2}
	\begin{bmatrix} 
		1 + e^{-t/\tau} & 1 - e^{-t/\tau} & 0 & 0 \\
		1 - e^{-t/\tau} & 1 + e^{-t/\tau} & 0 & 0 \\
		0 & 0 & 1 + e^{-t/\tau} & 1 - e^{-t/\tau} \\
		0 & 0 & 1 - e^{-t/\tau} & 1 + e^{-t/\tau}
	\end{bmatrix} $
	\\
	\hline
	Damping
	& $ \Phi_t[\rho] =
	\begin{bmatrix}
	1 - e^{-t/\tau} \rho_{11} & e^{-t/2\tau} \rho_{01} \\
	e^{-t/2\tau} \rho_{10} & e^{-t/\tau} \rho_{11}
	\end{bmatrix} $
	&
	$\begin{bmatrix}
	a  & b & c & c \\
	b  &  a  & c & c \\
	d  & d  & e  & f \\
	d & d  & f & e
	\end{bmatrix}$,
	
	where
	$a=\alpha+\frac{e^{-t/2\tau}}{2}, b =\alpha-\frac{e^{-t/2\tau}}{2}, c=\alpha-\frac{e^{-t/\tau}}{2} ,$ \\  
	& & 
	$d= -\alpha+\frac{1}{2},	 
	e=\alpha+\frac{e^{-t/\tau}}{2\sqrt{3}}+\frac{e^{-t/2\tau}}{2}-\frac{1}{2\sqrt{3}} ,$ \\
	& &
	$f=\alpha+\frac{e^{-t/\tau}}{2\sqrt{3}}-\frac{e^{-t/2\tau}}{2}-\frac{1}{2\sqrt{3}} ,$ and $ \alpha=\frac{e^{-t/\tau}}{4}-\frac{e^{-t/\tau}}{4\sqrt{3}}+\frac{1}{4}+\frac{1}{4\sqrt{3}}$\\
	\hline

	Rotation around $x$-axis
	& $ \Phi_t[\rho] = e^{- \frac{i \omega t}{2} \sigma^{(1)}} \rho e^{\frac{i \omega t}{2} \sigma^{(1)}} $
	& $\frac{1}{2}
	\begin{bmatrix}
 	2\cos^2(\omega t/2) & -\sin(\omega t) & \sin(\omega t) & 2\sin^2(\omega t/2) \\
	 \sin(\omega t) & 2\cos^2(\omega t/2) & 2\sin^2(\omega t/2) & -\sin(\omega t) \\
	 -\sin(\omega t) & 2\sin^2(\omega t/2) & 2\cos^2(\omega t/2 & \sin(\omega t) \\
	 2\sin^2(\omega t/2) & \sin(\omega t) & -\sin(\omega t) & 2\cos^2(\omega t/2)
	 \end{bmatrix}$
	\\
	\hline
	Rotation around $y$-axis
	& $ \Phi_t[\rho] = e^{- \frac{i \omega t}{2} \sigma^{(2)}} \rho e^{\frac{i \omega t}{2} \sigma^{(2)}} $
	& $\frac{1}{2}
	\begin{bmatrix}
	2\cos^2(\omega t/2 & -\sin(\omega t) & 2\sin^2(\omega t/2) & \sin(\omega t) \\
	\sin(\omega t) & 2\cos^2(\omega t/2 & -\sin(\omega t) & 2\sin^2(\omega t/2) \\
	2\sin^2(\omega t/2) & \sin(\omega t) & 2\cos^2(\omega t/2 & -\sin(\omega t) \\
	-\sin(\omega t) & 2\sin^2(\omega t/2) & \sin(\omega t) & 2\cos^2(\omega t/2
	\end{bmatrix}$
	\\
	\hline
	Rotation around $z$-axis
	& $ \Phi_t[\rho] = e^{- \frac{i \omega t}{2} \sigma^{(3)}} \rho e^{\frac{i \omega t}{2} \sigma^{(3)}} $
	& $\frac{1}{2}
	\begin{bmatrix}
	2\cos^2(\omega t/2 & 2\sin^2(\omega t/2) & \sin(\omega t) & -\sin(\omega t) \\
	2\sin^2(\omega t/2) & 2\cos^2(\omega t/2 & -\sin(\omega t) & \sin(\omega t) \\
	-\sin(\omega t) & \sin(\omega t) & 2\cos^2(\omega t/2 & 2\sin^2(\omega t/2) \\
	\sin(\omega t) & -\sin(\omega t) & 2\sin^2(\omega t/2) & 2\cos^2(\omega t/2
	\end{bmatrix} $ \\
	\hline
	\end{tabular}
	\label{tbl:generators-matrices}
	\end{table*}	
	
\begin{figure*}
			\centering
		\includegraphics[width=1.02\linewidth]{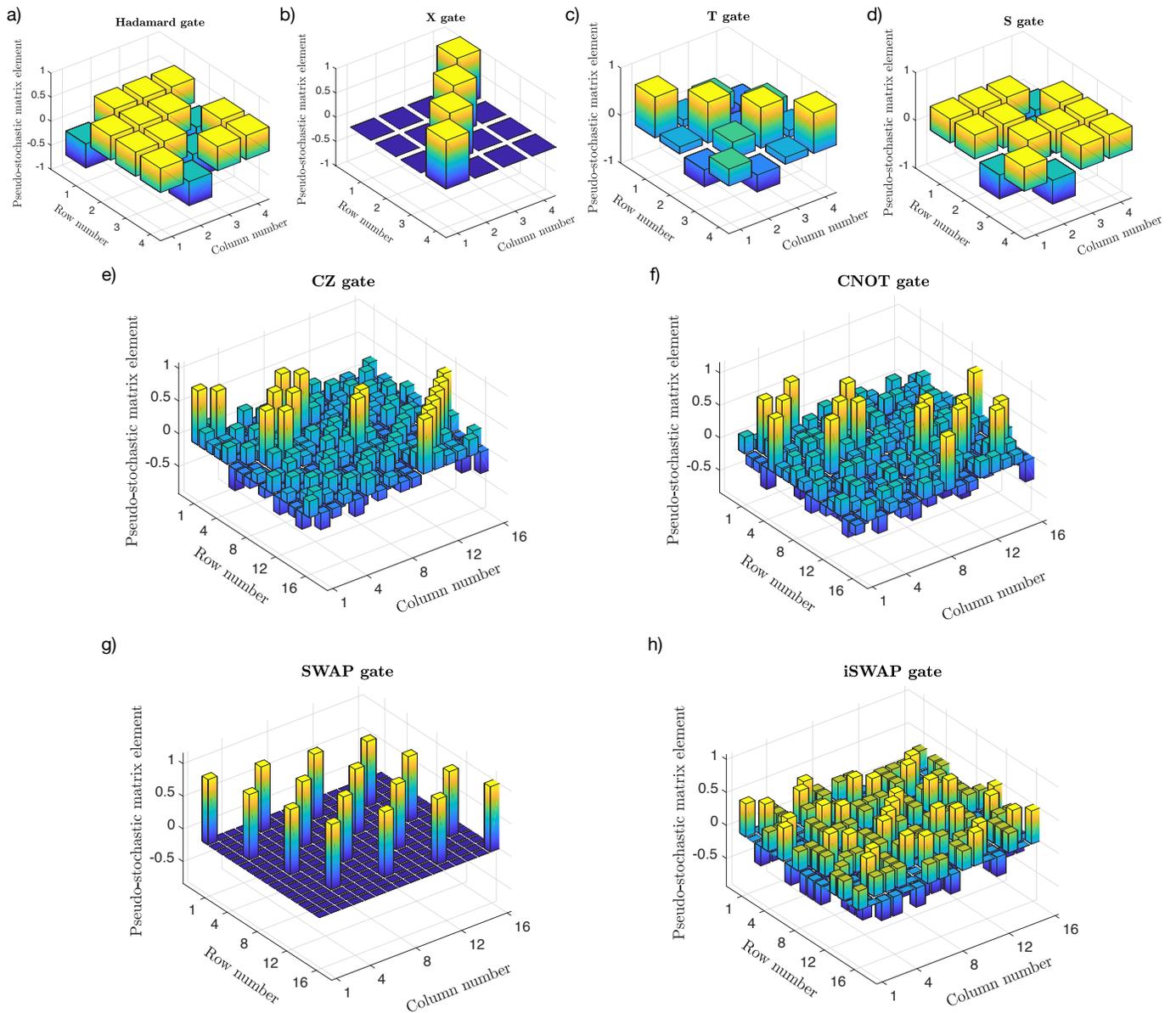}
		\caption{Pseudostochastic matrices of a single- and two-qubit gates: (a) Hadamard gate, b) X-gate, c) T-gate, d) S-gate, e) CZ-gate, f) CNOT-gate, g) SWAP gate, and h) iSWAP gate.}
		\label{fig:gates}
	\end{figure*}

	We also demonstrate pseudostochastic matrices for some single- and two-qubit gates in Fig.~\ref{fig:gates}. 
	Pseudostochastic matrices of singe-qubit gates are obtained with SIC-POVM $E^{\rm sym}$ given in Eq.~\eqref{eq:SIC-POVM}.
	For constructing probability representation of two-qubit gates, the MIC-POVM $E^{\rm sym}\otimes E^{\rm sym}$ is used.
	One can observe an appearance of negative elements in pseudostochastic matrices of Hadamard gate, $T$ gate, $S=T^{2}$ gate; $CZ$ gate, CNOT ($CX$) gate, and iSWAP gate.
	We note that negative elements are absent for Pauli-X gate and SWAP gate.

\end{document}